\documentclass[accepted=2021-11-21,a4paper]{quantumarticle}
\pdfoutput=1 

\usepackage{graphicx}
\usepackage{amsmath} 
\usepackage{dcolumn}
\usepackage{bm}

\usepackage{amssymb} 
\usepackage[utf8x]{inputenc} 
\usepackage{braket}
\usepackage{ifthen} 
\usepackage{dsfont} 
\usepackage{multirow} 
\usepackage{enumitem}   
\usepackage{wrapfig,array}
\usepackage{scalerel} 
\usepackage{bbm}
\usepackage{pgf} 
\usepackage{appendix}
\usepackage{relsize}
\usepackage{textcomp}
\usepackage{color, colortbl} 
\usepackage{hyperref}
\usepackage{url}
\usepackage[subtle,mathspacing=normal,tracking=normal]{savetrees}

\usepackage{xkeyval,etoolbox,geometry,xcolor,fancyhdr,tikz,ltxgrid,ltxcmds,lmodern}

\usepackage[numbers,sort&compress]{natbib}

\def\eps{\epsilon}

\def\varth{\vartheta}

\def\vf{\varphi}

\def\dZ{\mathbb{Z}}
\def\dR{\mathbb{R}}

\def\sC{\mathcal{C}}

\DeclareRobustCommand*{\sH}{\mathcal{H}}
\def\sI{\mathcal{I}}
\def\sJ{\mathcal{J}}
\def\sK{\mathcal{K}}
\def\sL{\mathcal{L}}

\def\sQ{\mathcal{Q}}
\def\sR{\mathcal{R}}
\def\cS{\mathcal{S}}
\def\sT{\mathcal{T}}

\def\sV{\mathcal{V}}

\def\sX{\mathcal{X}}

\def\sZ{\mathcal{Z}}

\def\eff{\mathrm{eff}}
\def\sV{\mathcal{V}}
\def\uno{\mathbbm{1}}

\newcolumntype{P}[1]{>{\centering\arraybackslash}p{#1}} 

\newcommand{\cdc}[2]{c^\dagger_{#1} c^{\vphantom{\dagger}}_{#2}}

\renewcommand{\k}[1]{\ket{#1}}

\newcommand{\kb}[2]{\ket{#1}\bra{#2}} 

\newcommand{\diag}{\mathop{\mathrm{diag}}}

\def\rdos{\sqrt{2}}

\newcommand*{\rhomb}[1][]{%
  \pgfpicture\pgfsetroundjoin
    \pgftransformxslant{.6}%
    \pgfpathrectangle{\pgfpointorigin}{\pgfpoint{.60em}{.65em}}%
    \pgfusepath{stroke}%
  \endpgfpicture}

\newcommand{\subrhomb}{\scalebox{0.5}{\rhomb}}

\newcommand{\hgls}{\mathbin{\rotatebox[origin=c]{90}{\scalebox{1.5}{$\Join$}}}}
\newcommand{\subhgls}{\mathbin{\rotatebox[origin=c]{90}{\scalebox{0.7}{$\Join$}}}}

\colorlet{Gray}{lightgray!40!}

\newcolumntype{B}{>{\columncolor[rgb]{0.9,0.9,0.9}}c}

\begin{document}


\title{Tunable zero modes and quantum interferences in flat-band topological insulators}

\author{Juan Zurita}
\affiliation{
Instituto de Ciencia de Materiales de Madrid (CSIC), Cantoblanco, E-28049 Madrid, Spain
}%
\affiliation{
 Universidad Complutense de Madrid, E-28040 Madrid, Spain
}%

\author{Charles Creffield}%
\affiliation{
 Universidad Complutense de Madrid, E-28040 Madrid, Spain
}%


\author{Gloria Platero}
\email[]{gplatero@icmm.csic.es}
\affiliation{
Instituto de Ciencia de Materiales de Madrid (CSIC), Cantoblanco, E-28049 Madrid, Spain
}%


\begin{abstract}

We investigate the interplay between Aharonov-Bohm (AB) caging and topological protection in a family of quasi-one-dimensional topological insulators, which we term CSSH ladders. Hybrids of the Creutz ladder and the SSH chain, they present a regime with completely flat bands, and a rich topological phase diagram, with several kinds of protected zero modes. 
These are reminiscent of the Creutz ladder edge states in some cases, and of the SSH chain edge states in others. Furthermore, their high degree of tunability, and the fact that they remain topologically protected even in small systems in the rungless case, due to AB caging, make them suitable for quantum information purposes.
One of the ladders can belong to the BDI, AIII and D symmetry classes depending on its parameters, the latter being unusual in a non-superconducting model.
Two of the models can also harbor topological end modes which do not follow the usual bulk-boundary correspondence, and are instead related to a Chern number.
Finally, we propose some experimental setups to implement the CSSH ladders with current technology, focusing on the photonic lattice case.
\end{abstract}

\maketitle

\section{\label{sec:Intro}Introduction}

Topological materials have a gapped bulk, which can be insulating or superconducting, and a metallic surface spectrum. For one-dimensional (1D) materials, this implies the existence of topological zero modes, robust end states pinned at zero energy and protected by symmetries of the Hamiltonian. Depending on these symmetries, topological materials can be classified into ten topological phases, five of which can show topological phases in 1D \cite{Altland1997}.

The robustness of their topological end modes can serve as a way to prevent errors in quantum information protocols. For instance, the well-known SSH chain end modes can be used as a way to implement a robust quantum state transfer protocol \cite{Lang2017, Mei2018, Qi2020,Estarellas2017,Boross2019}, or a quantum memory resistant to off-diagonal disorder \cite{Estarellas2017}.
Ladder models are quasi-1D lattice models that consist of two chains of sites coupled by tunneling amplitudes or interactions. By tuning the different parameters in ladder models, several BDI \cite{Tovmasyan2013,Li2015,Padavic2018,Velasco2019}, AIII \cite{Piga2017, Sun2017, Velasco2019,Kuno2020,VanDalum2021} and CII \cite{Liu2019,Gholizadeh2018} topological insulators (TIs) have been found.
The best-known topological ladder model is probably the Creutz ladder \cite{Creutz1999}, a BDI topological insulator that also shows flat bands in a certain regime, where particles experience a self-localization phenomenon termed ``Aharonov-Bohm (AB) caging’’ due to a magnetic flux. This effect has previously been studied in other 1D and 2D lattices, like the rhombus chain \cite{Vidal2000,Vidal2002,Creffield2010,Pelegri2019} and the $\sT_3$ and $\sT_4$ lattices \cite{Vidal1998,Vidal2001}, and has been observed experimentally in several setups \cite{Abilio1999,Naud2001,Shinohara2002,Mukherjee2018,Kremer2020,Jorg2020}.

\begin{figure}[!htbp] 
    \includegraphics[width=\linewidth,trim=10 10 10 3,clip]{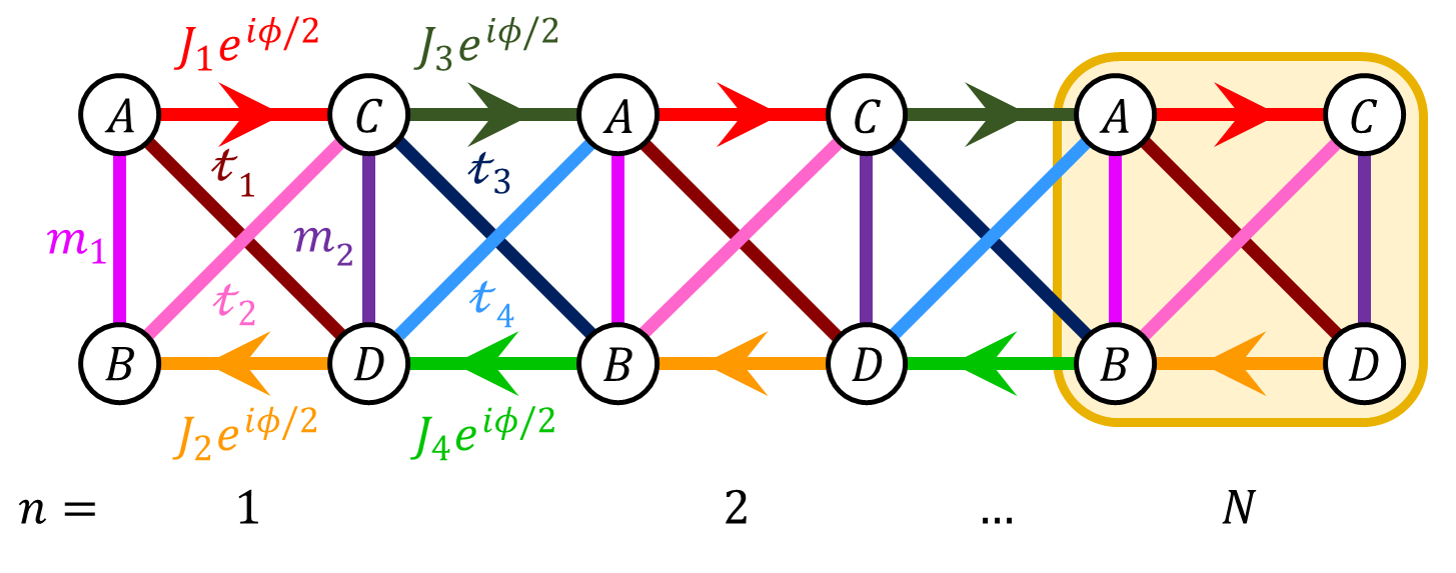}
    \caption{CSSH ladder diagram in the general case. The model is determined by its four horizontal ($J_i$), and four diagonal ($t_i$) tunneling amplitudes, its two vertical tunneling amplitudes $m_1,m_2$ (which we will consider equal) and the magnetic flux $\phi$. A unit cell is shaded in orange.}
    \label{fig:Gen}
\end{figure}

Flat-band models have attracted a great deal of research in recent years, in part because of the discovery of magic-angle twisted bilayer graphene \cite{Cao2018}. The Creutz ladder and other quasi-1D flat-band models like the rhombus chain can serve as toy models to explore the interplay of localization dynamics with interactions or topology \cite{Creffield2010, Tovmasyan2013, Takayoshi2013, Mondaini2018, Zurita2019,Pelegri2019}. Other studied variants of the Creutz ladder include additional Peierls phases or on-site energies \cite{Hugel2014,Sun2017,Kuno2020,Piga2017}, Floquet driving \cite{Zhou2019,Molignini2020}, spin-$1/2$ particles \cite{Liu2019a,Zhou2019,Gholizadeh2018} or superconductivity \cite{Sticlet2014,Mondaini2018}.

\begin{figure}[!htbp] 
    \includegraphics[width=\linewidth,trim=10 10 10 3,clip]{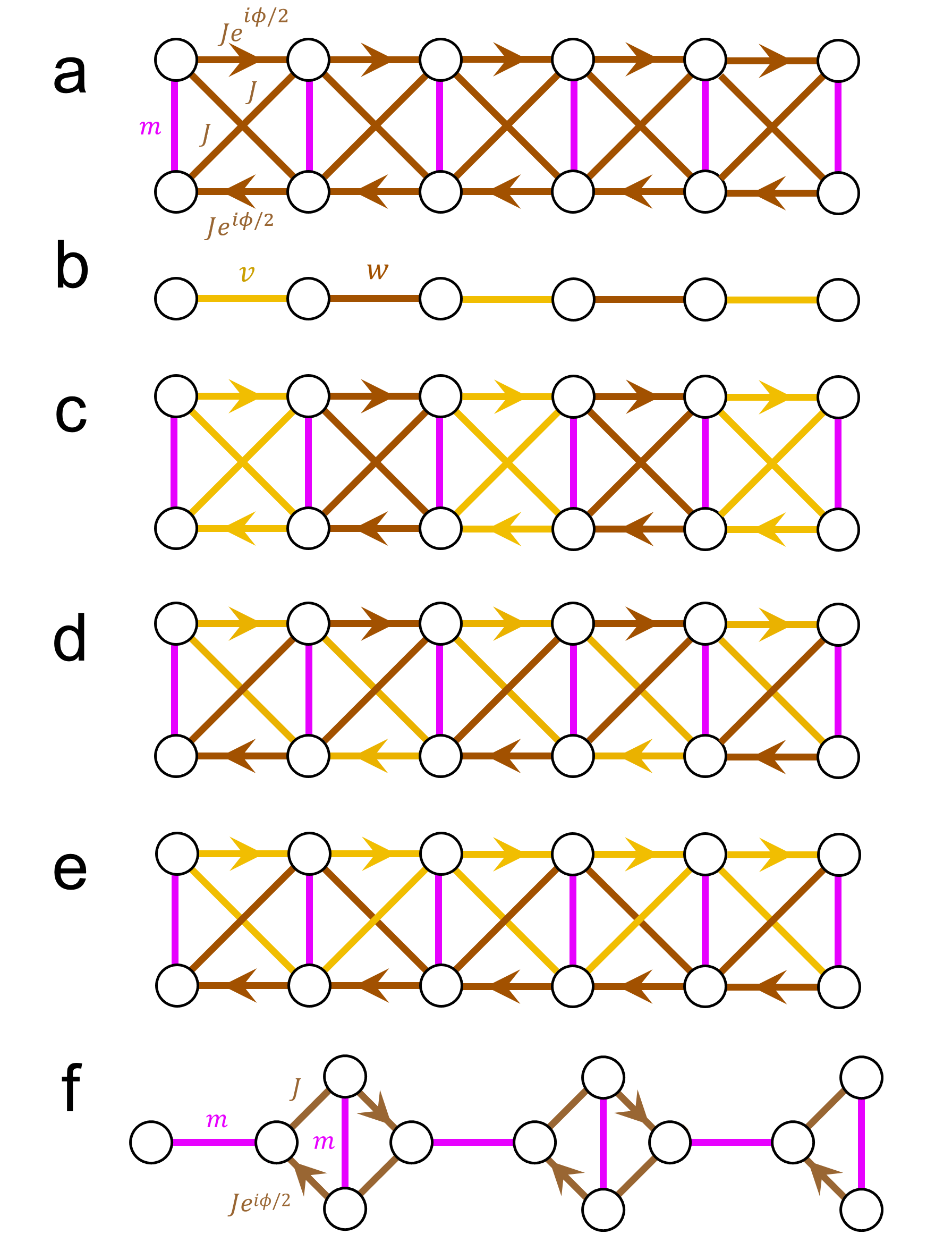}
    \caption{(a) The Creutz ladder, obtained from the general case (Fig. \ref{fig:Gen}) when $J_i=t_i=J$ (in dark brown), and $m_1=m_2$ (in purple). (b) The SSH chain, with tunneling amplitudes $v$ and $w$. (c) The hourglass CSSH ladder, with $J_3=J_4=t_3=t_4=J$, $J_1=J_2=t_1=t_2=J'$. The darker (lighter) links represent $J$-valued ($J'$-valued) amplitudes. (d) The rhomboid CSSH ladder, with $J_2=J_3=t_2=t_4=J$, $J_1=J_4=t_1=t_3=J'$. (e) The delta CSSH ladder, with $J_2=J_4=t_2=t_3=J$, $J_1=J_3=t_1=t_4=J'$. (f) The limit of the rhomboid CSSH ladder when $J'=0$ is the orthogonal dimer chain, which is also a topological insulator.}
    \label{fig:CSSH}
\end{figure}

In this work, we propose and investigate a family of quasi-1D topological insulators related to the Creutz ladder and the SSH chain, which we term Creutz-Su-Schrieffer-Heeger (CSSH) ladders [see Figs. \ref{fig:Gen} and \ref{fig:CSSH}]. They present a regime where all four bands are flat due to AB caging. This phenomenon also localizes the edge states in one or two sites whenever the vertical links of the ladders vanish. This prevents the overlap of the left and right states, and allows them to be degenerate at zero energy even in small systems. This unusual interplay between quantum interference and topology, inherited from the Creutz ladder, enhances the topological protection of the states, which is preserved even in the one-plaquette limit of the ladders. A remarkable result of our work is that, in the CSSH ladders, the wavefunction of these end states can be tuned, using the parameters of the model, to take the form of almost any state in the Bloch sphere spanned by the two end sites. This can be used to prepare the Creutz ladder end states starting from a particle in a single site, or to control a particle in a topologically protected way even in small systems, something that can be useful for quantum information purposes.

Interestingly, one of the models, the hourglass CSSH ladder, can harbor topological zero modes reminiscent of both the Creutz ladder and the SSH chain end states. Another one of them, the delta CSSH ladder, has topological phases in the BDI, AIII and D classes, such that these three classes can be explored by changing the parameters of the system, without closing the gap or losing the protection of the states. This is something remarkable, because models in the D class in 1D are usually topological superconductors \cite{Budich2013,Lin2017}. 
In the rhomboid and delta CSSH ladders, we also report the existence non-protected topological end modes which do not follow the usual bulk-boundary correspondence. They can instead be related to a Chern number, even though they appear in a quasi-1D model. These modes are very similar to those found in the trimer chain \cite{Martinez2019}.
The mentioned features make the CSSH ladders versatile and exotic models, which can also be realized experimentally in several bosonic and fermionic platforms with current technology.

In Section \ref{sec:CSSH}, we present the Hamiltonian of the CSSH ladders. Then, in section \ref{sec:simple}, we review the topology of the SSH chain and the Creutz ladder, including an original full description of the symmetries of the latter. In Section \ref{sec:topo}, we discuss the topological phase diagrams of the models, their spectra, symmetries and zero modes. In Section \ref{sec:exp} we propose some experimental implementations of the models, focusing on the photonic lattice case. Finally, in Section \ref{sec:conclusion}, we present our conclusions.

\phantom{\footnotesize.}

\section{The CSSH ladder models}\label{sec:CSSH}


The Creutz ladder \cite{Creutz1999,Piga2017,Zurita2019}, a quasi-1D topological insulator, is formed by two chains of sites, connected by vertical, horizontal, and diagonal links [see Fig. \ref{fig:CSSH} (a)]. A magnetic field is applied in the perpendicular direction.
On the other hand, the Su-Schrieffer-Heeger (SSH) chain \cite{Asboth2015,Bello2016}, the simplest 1D TI, consists of a chain of sites connected by alternate hopping amplitudes [see Fig. \ref{fig:CSSH} (b)]. 
Both systems are chiral TIs for some values of their parameters (see Section \ref{sec:simple}).

In this work, we consider a family of models closely related to the Creutz ladder, which will also include the alternating hopping amplitudes of the SSH model. Therefore, we will refer to them as CSSH ladders. 
To label each site, we will use the unit cell number $n=1,\ldots,N$ and the internal coordinate $\alpha=A,B,C,D$, as indicated in Fig. \ref{fig:Gen}. We will also use the nomenclature $\overline A = B$, $\overline B = A$, $\overline C = D$, $\overline D = C$. 
The general Hamiltonian for these models is (see Fig. \ref{fig:Gen}):
\begin{widetext}
\begin{align}
    &\sH_{C\!S\!S\!H} = -\sum^{\left\lfloor L/2 \right\rfloor}_{n=1} \left(
    J_1^+ \cdc{n,C}{n,A} +
    J_2^- \cdc{n,D}{n,B} + 
    J_3^+ \cdc{n+1,A}{n,C} +
    J_4^- \cdc{n+1,B}{n,D} +\right.\nonumber\\
    &\left.+t_1 \cdc{n,D}{n,A} +
    t_2 \cdc{n,C}{n,B} +
    t_3 \cdc{n+1,B}{n,C} +
    t_4 \cdc{n+1,A}{n,D} +
    m_2 \cdc{n,C}{n,D} + h.c. \right) +\nonumber\\
    &  -\sum^{\left\lceil L/2 \right\rceil}_{n=1} \left( m_1 \cdc{n,A}{n,B}  +h.c.\right), \label{eq:CSSH}
\end{align}
where $\lfloor \cdot \rfloor$ and $\lceil \cdot \rceil$ are the floor and ceiling functions, and $J_i^\pm = J_i e^{\pm i\phi/2}$ for $i=1,2,3,4$. The operator $c^{(\dagger)}_{n,\alpha}$ destroys (creates) a particle in site $n,\alpha$. Our work is applicable to both bosons and spinless (or spin-polarized) fermions. The upper limits of the sums have been chosen to take into account the possibility of a finite system with an odd length $L$, in which the last unit cell is not complete. When not explicitly noted otherwise, we will assume that $L$ is even, and then $\left\lfloor L/2 \right\rfloor=\left\lceil L/2 \right\rceil=N$, the total number of unit cells. We consider the same phase shift $\phi$ in every horizontal link, implementing a magnetic flux of $\Phi = \hbar\phi/q$ through each square plaquette, where $q$ is the charge of the particle.

We will also fix the vertical hopping amplitudes as equal, $m_1 = m_2 = m$. This leaves 10 degrees of freedom, $\{J_i,t_i,\phi,m\}_{i=1}^4$.
Under periodic boundary conditions, this Hamiltonian can be rewritten as $\sH_{C\!S\!S\!H} = \sum_{k\in BZ}\ket{k}\bra{k}\otimes \sH(k)$, where the first operator acts on the spatial coordinate (that is, the unit cell number $n$), while the $4\times 4$ matrix $\sH(k)$ acts on the internal degree of freedom $\alpha$. The quasimomentum $k$ takes values along the first Brillouin zone. The bulk Hamiltonian has the form: 
\begin{equation} \label{eq:Hk}
    \sH(k) = - \begin{pmatrix}
    0 & m & J_1^- + J_3^+e^{-ik} & t_1 + t_4e^{-ik}\\
    m & 0 & t_2+t_3e^{-ik} & J_2^+ + J_4^-e^{-ik}\\
    J_1^+ + J_3^-e^{ik} & t_2+t_3e^{ik} & 0 & m\\
    t_1+t_4e^{ik} & J_2^- + J_4^+e^{ik} & m & 0
    \end{pmatrix},
\end{equation}
and will allow us to probe the topology of the systems in Section \ref{sec:topo}.
\end{widetext}

Out of all possible systems, we analyze particular cases in which the horizontal and diagonal hoppings, $J_i$ and $t_i$, take two different values, $J$ and $J'$, in the spirit of the SSH model. In addition, we will focus our attention on cases where all energy bands are non-dispersive due to AB caging when $\phi=\pi$ and $m=0$. The CSSH models will have, in general, 4 different bands, given that their unit cell consists of 4 sites. In addition to the Creutz ladder [Fig. \ref{fig:CSSH} (a)], where $J_i=t_i=J$ $\forall i$, three different kinds of CSSH ladders that exhibit this behavior are:
\begin{itemize}
	\item The hourglass ($\hgls$) CSSH ladder, with $J_3=J_4=t_3=t_4=J$, $J_1=J_2=t_1=t_2=J'$ [Fig. \ref{fig:CSSH} (c)]. 
	\item The rhomboid ($\rhomb$) CSSH ladder, with $J_2=J_3=t_2=t_4=J$, $J_1=J_4=t_1=t_3=J'$ [Fig. \ref{fig:CSSH} (d)]. 
	\item The delta ($\Delta$) CSSH ladder, with $J_2=J_4=t_2=t_3=J$, $J_1=J_3=t_1=t_4=J'$ [Fig. \ref{fig:CSSH} (e)]. 
\end{itemize}
The name for each kind of ladder comes from the shape formed by the links of the same value. We will parametrize $J$ and $J'$ in the following way:
\begin{equation}
	J = \xi \sin^2 \theta; \hspace{20pt} J' = \xi \cos^2 \theta,
\end{equation} 
with $0\leq\theta < \pi/2$. Thus, the Hamiltonian for any of the CSSH ladders is completely determined by the hopping amplitudes $\xi = J + J'$ and $m$, the angle $\theta$ and the magnetic flux phase $\phi$. Unless specified otherwise\footnote{Some properties of the models, like the winding number or the chirality of the zero modes, are periodic in $\phi$ with period $4\pi$, not $2\pi$. This is related to the fact that the smallest closed loop in the $m\ne 0$ model is a triangle with a flux of $\phi/2$. However, most features explored in this work have a period of $2\pi$.}, all values of $\phi$ are to be taken modulo $2\pi$.  From this point on, we set $\hbar = 1$ and we consider $\xi=1$, which sets the energy scale of the system but has no other effect in the dynamics. As we will see below, the CSSH ladders inherit some features from the Creutz ladder, and some others from the SSH chain.

These three models have not been previously studied in the literature, although they bear some relation to other quasi-1D insulators. The $\rhomb$-CSSH ladder, when either $J$ or $J'$ are zero, reduces to the orthogonal dimer chain [Fig. \ref{fig:CSSH} (f)], a topological model that is usually studied as a chain of interacting spins \cite{Ivanov1997,Richter1998,Canova2004,Paulinelli2013,Verkholyak2013,Leykam2018,Nandy2019,Rojas2019,Galisova2021}. The $\rhomb$-CSSH ladder model connects the topological phases of the orthogonal dimer chain and of the $\pi$-flux Creutz ladder in an adiabatic way, without losing the topological protection of the edge states at any point\footnote{In this regard, this model is similar to the $\alpha$-$\sT_3$ lattice, a non-topological 2D model with a flat band and two different tunneling amplitudes, which interpolates between the honeycomb lattice and the $\sT_3$ lattice \cite{Raoux2014,Illes2016,Illes2017}.}.

The spin-$1/2$ SSH chain with spin-orbit interaction \cite{Yan2014,Han2014,Jiang2015,Bahari2020} is similar to the hourglass CSSH ladder in a certain regime, except for the distribution of magnetic fluxes. The hourglass-CSSH ladder with $\phi=0$, in its trivial phase, also has the same geometry as certain $\text{Cu}_2\text{Te}_2\text{O}_5\text{X}_2$ compounds with X$=$Cl, Br \cite{Volkova2017}; and has been analyzed as a chain of interacting spins \cite{Brenig2001,Volkova2017}. Other models similar to the CSSH ladders, but with some missing tunneling amplitudes, have been studied, like the SSH ladder \cite{Ootomo2000,Zhang2017a,Padavic2018,Nersesyan2020,Jangjan2020}, the SSH bowtie ladder \cite{Velasco2019,Ryu2020} or the chequerboard chain \cite{Drescher2017}. A thorough exploration of other topological ladders can be found in \cite{Velasco2019}.

\section{Topology in simple models}\label{sec:simple}
\begin{figure*}[!htbp] 
    \includegraphics[width=\linewidth,trim=10 10 10 10,clip]{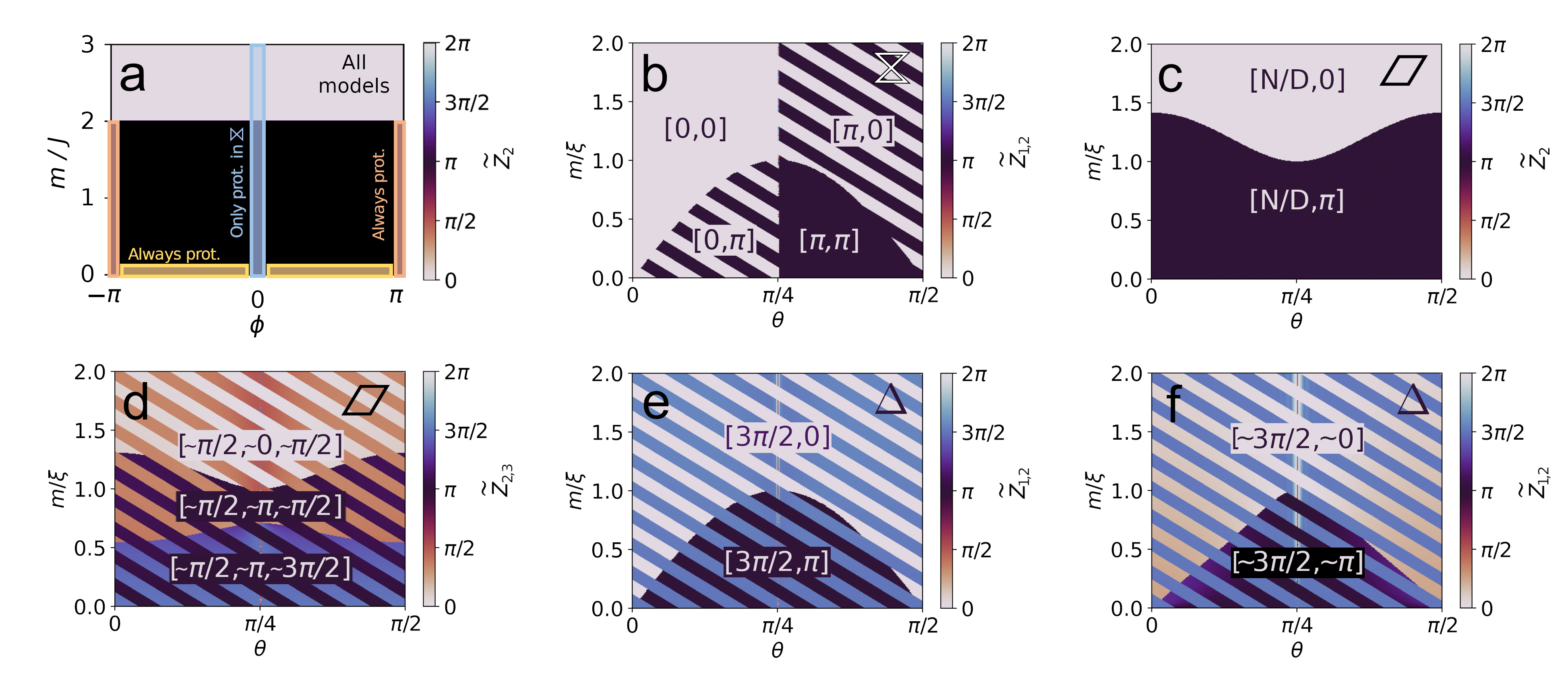}
    \caption{Topological phase diagrams (TPDs). Some figures are striped to show the value of two invariants, which are $[\tilde\sZ_1,\tilde\sZ_2]$ unless specified otherwise. (a) TPD for all models in the Creutz limit, $\theta = \pi/4$, represented by its Zak phase $\sZ$, which takes the values of 0 (trivial) or $\pi$ (topological). Overlapped on the diagram, we mark the three regimes in the $(\phi,m)$ plane where the end states can be protected, for at least some value of $\theta$: $\phi=\pi$ (orange) and $m=0$ (yellow) in all systems, and $\phi=0$ in the $\hgls$-CSSH case (blue). For other values of $\phi$, the Zak phase is quantized but the states are not protected if $m>0$. 
    (b) $\hgls$-CSSH ladder TPD when $\phi=\pi$. The phase diagram looks identical when $\phi\ne 0,\pi$, but the states are not protected there when $m\ne 0$. 
   (c) $\rhomb$-CSSH ladder TPD when $\phi=\pi$. $\tilde\sZ_1$ is not defined (N/D), because the bands touch [see Fig. \ref{fig:3Dspectra} (c)], so only $\tilde\sZ_2$ is pictured.
   (d) $\rhomb$-CSSH ladder TPD for $\phi=\pi/2$. The absence of inversion symmetry means that the Zak phase is not quantized to only 0 or $\pi$. Additionally, $\tilde\sZ_1 \ne \tilde\sZ_3$ here, so we identify the phases by  $[\tilde\sZ_1,\tilde\sZ_2,\tilde\sZ_3]$, see the main text for details. In this case, the diagram is striped to show $\tilde\sZ_2$ and $\tilde\sZ_3$.
   Three phases arise, where the invariants oscillate around a given value, something we indicate with a $\sim$ symbol. The model belongs to a topological symmetry class (BDI) only for $m=0$, where the second gap has protected zero modes.
   (e) $\Delta$-CSSH ladder TPD for $\phi=\pi$. The first invariant has a constant value of $3\pi/2$, except at the Creutz limit $\theta=\pi/4$, where it becomes undefined. 
   (f) $\Delta$-CSSH laddder TPD for $\phi=\pi/2$. It is similar to the $\pi$-flux case, but the values of the invariants now oscillate around the indicated values. In the $m=0$ line (AIII case), the second gap is topological.
    All results are obtained from the bulk Hamiltonian $\sH(k)$, following the formula \eqref{eq:Zak} using 100 values of $k$. All CSSH ladders have $\xi=1$.}
    \label{fig:Zak}
\end{figure*}

The three main symmetries that are usually considered in the context of topology are particle-hole (PH) symmetry ($\sC$) and time-reversal (TR) symmetry ($\sT$), which are both antiunitary, and chiral or sublattice symmetry ($\sX=\sC \sT$), which is unitary.
They satisfy:
\begin{align}
    &\sC \sH \sC^{-1} = -\sH \label{eq:sC}\\
    &\sT \sH \sT^{-1} = \sH \label{eq:sT}\\
    &\sX \sH \sX^{-1} = -\sH \label{eq:sX}.
\end{align}
Their presence and type define the well-known symmetry classes in the periodic table of topological insulators \cite{Altland1997,Schnyder2008,Kitaev2009}, some of which can have topological phases, depending on their spatial dimension. The five topological classes in 1D can be consulted in Table \ref{tab:clas}.

In this paper, we follow the approach of considering the requirements (\ref{eq:sC}-\ref{eq:sX}) as the defining property of the three symmetries, which is a useful point of view for the analysis of topological insulators and has been used extensively in the literature \cite{Budich2013,Stanescu2016,Piga2017,Gholizadeh2018,Alaeian2019,Leonforte2021}. However, it is important to note that the particle-hole symmetries we discuss do not have the same implications as in a topological superconductor. Specifically, it does not imply the presence of Majorana end modes in the topological phases. The unit cells in the models we discuss are composed of more than one particle mode, and so their end modes are never formed by half a fermionic mode \cite{Jackiw2012,Budich2013,Stanescu2016,Herviou2017}.

\begin{table}[!htbp]
\begin{tabular}{P{0.1\textwidth}P{0.001\textwidth}P{0.03\textwidth}P{0.003\textwidth}P{0.01\textwidth}P{0.01\textwidth}P{0.015\textwidth}P{0.025\textwidth}P{0.01\textwidth}P{0.003\textwidth}P{0.003\textwidth}}\hline 
    Class   & \!\!A\!\! &\!\! AIII\!\! & \!\!AI\!\! & \!\!BDI\!\! &\!\! D\!\!     & \!\!DIII\!\! &\!\! AII\!\! & \!\!CII\!\! & \!\!C\!\! & \!\!CI\!\! \\\hline
     1D T. Inv.  &   &$\dZ$\!\!  &  &  $\dZ$\!\!&   $\dZ_2$\!\!&\!\! $\dZ_2$\!\!&    &\!\!$2\dZ$\!\!&   &    \\ \hline
\end{tabular}
\caption{Altland-Zirnbauer symmetry classes for 1D systems and their topological invariants \cite{Altland1997}. The delta CSSH ladder spans three of the five topological classes, AIII, BDI and D, while the other CSSH ladders and the Creutz ladder can be BDI or AIII.} \label{tab:clas}
\end{table}

In this section, we will review the relevant symmetries of the SSH chain and Creutz ladder, including the hidden symmetries in the Creutz ladder, essential for the protection of the edge states. Then, we will describe the symmetries and topology of the three CSSH ladder models proposed in this work. All the cases we consider are represented in Table \ref{tab:syms}, in which the symmetry classes and invariants of the different models are detailed.

\subsection{The SSH chain}
The SSH chain \cite{Su1979} [Fig. \ref{fig:CSSH} (b)] is a 1D chain with two sites per unit cell, $I$ and $I\!I$, with a intracell tunneling amplitude of $v$, and an intercell amplitude of $w$, in which particles move around. Its Hamiltonian is:
\begin{equation}\label{eq:SSH}
    \sH_{S\!S\!H} \!= \!-\!\sum_{x} \left(v c^\dagger_{x,I\!I} c_{x,I} + w c^\dagger_{x+1,I} c_{x,I\!I}    + h.c.\right)\!,
\end{equation}
where $c^{(\dagger)}_{x,\sigma}$ destroys (creates) a particle in unit cell number $x=1,\ldots,L/2$ and sublattice $\sigma=I,I\!I$.

This Hamiltonian has particle-hole ($\sC^{(2)}_S=\sigma_z \sK$), time-reversal ($\sT^{(2)}_S=\sK$) and chiral ($\sX^{(2)}_S=\sigma_z$) symmetries, where $\sK$ is complex conjugation. The matrices act on the internal degree of freedom, the subindex $S$ refers to the SSH chain, and the superindex indicates the matrices are $2\times 2$.
The model belongs to class BDI \cite{Asboth2015,Altland1997}, and so it can be characterized by its Zak phase (for a more detailed discussion, please refer to Appendix \ref{ap:zak}).
The Zak phase of a given band $i$ is given by\footnote{As detailed in Appendix \ref{ap:zak}, there are two slightly different quantities that have been called ``Zak phase'' in the literature. We follow the convention in \cite{Asboth2015}, which is tied to the choice of the so-called Basis I for tight-binding models \cite{Bena2009}.}:
\begin{equation} \label{eq:Zak}
	\sZ_i = i \oint_{BZ} \bra{u_k^i} \partial_k \ket{u_k^i} dk,
\end{equation}
where $\{\ket{u_k^i}\}$ is the $i$-th eigenstate of the bulk Hamiltonian $\sH(k)$ of the system \cite{Zak1989}.

The internal states of the SSH chain $\k{I}=(1,0)$ and $\k{I\!I}=(0,1)$ have positive and negative chirality, respectively, corresponding to the two sublattices. The bulk states have equal support on both sublattices, while the topological edge states, when present, only have support on one of them.
This is a general phenomenon in chiral-symmetric topological phases.
The SSH chain is topological whenever $v<w$, giving rise to one chiral zero mode in each end of the chain, protected by the chiral and PH symmetries \cite{Ryu2002}.

\subsection{The Creutz ladder}

\subsubsection{ \texorpdfstring{$\phi=\pi$}{phi=pi} }

The Creutz ladder Hamiltonian can be obtained from equation \eqref{eq:CSSH} setting $J_i=t_i=J$. Its unit cell is formed by the upper and lower sites, $a$ and $b$, of each rung. The system is metallic when $m = 2J$ or $\phi = 0$ $\!\!\!\mod 2\pi$. When $\phi=\pi$, the system belongs to class BDI, with the symmetries $\sC^{(2)}_C = \sigma_z \sK$, $\sT^{(2)}_C = \sigma_x \sK$ and $\sX^{(2)}_C = i\sigma_y$ \cite{Sticlet2014,Hugel2014,Alaeian2019}, where the subindex $C$ refers to the Creutz ladder. This TR symmetry, as well as several others that we describe below, is non-conventional, but it behaves exactly like a usual TR symmetry for topological considerations, so we will consider it as such. For a more detailed discussion, see Appendix \ref{ap:syms}.

The topology of the model can be characterized by the Zak phase, which is $\pi$ if $m<2J$ and vanishes otherwise; except if $\phi=0,\pi \mod 2\pi$, where it is not defined because the system is metallic. These lines at constant $\phi$ in the phase diagram separate distinct topological phases, which can be distinguished by the chirality of their edge states or the sign of their winding number, $\nu=\pm 1$ \cite{Li2015,Sun2017}. One protected chiral edge state appears in each end of the ladder when the system is topological.

The chiral basis is now, in the $a,b$ basis, $(1,\mp i)/\sqrt{2}$, corresponding to the chirality of the left/right edge state. The edge states are protected by both the chiral and PH symmetries.

A unitary transformation \cite{Hugel2014} maps the $\phi=\pi$ ladder to the SSH chain, with $v=m$ and $w=2J$, as the topological phase transitions (TPTs) of each model suggest. For instance, this means that the rungless case ($m=0$), when AB caging completely localizes the topological edge states of the Creutz ladder in the first and last rungs, corresponds to the fully dimerized limit ($v=0$) in the SSH chain, where the edge states are localized at the first and last sites. This has important consequences, as we discuss below. The mapping also relates the chiral bases of the models. This illustrates a key difference between them: the two chiral states in the SSH chain coincide with its sites, while those in the Creutz ladder do not. While one of the chiral states (and, thus, one of the end states, if they are present) can be easily eliminated in the SSH chain due to open boundary conditions, this is impossible in the Creutz ladder. Thus, the topology of the Creutz ladder does not depend on the choice of unit cell, unlike that of the SSH chain.

\begin{table*}[htbp]
\small
\centering
\setlength\doublerulesep{8pt} 
\begin{tabular}{B ccc}
\hline
\cellcolor{Gray} \textbf{SSH}    &  \multicolumn{3}{c}{BDI: $\dZ$ [$\sC^{(2)}_S,\sT^{(2)}_S,\sX^{(2)}_S$]} \\ \hline \hline \rowcolor{Gray}[\tabcolsep]
\textbf{Creutz}                    & $\bm{\phi=0}$                                   & $\bm{\phi\ne 0,\pi}$                       & $\bm{\phi=\pi}$                                 \\ \hline
$\bm{m\ne 0}$ & Trivial & AI: NT [$\sT^{(2)}_C$] & BDI: $\dZ$ [$\sC^{(2)}_C\!,\sT^{(2)}_C\!,\sX^{(2)}_C$] \\ \hline
$\bm{m=0}$ & Trivial &  BDI$^a$: $\bm{\dZ}$ [$\tilde\sC_{SC},\sT^{(2)}_C\!, \tilde\sX_{S}$ ] & BDI: $\dZ$ [$\sC^{(2)}_C\!,\sT^{(2)}_C\!,\sX^{(2)}_C$]   \\ \hline \hline \rowcolor{Gray}[\tabcolsep]
$\hgls$\textbf{-CSSH }             & $\bm{\phi=0}$                                   & $\bm{\phi\ne 0,\pi}$                       & $\bm{\phi=\pi}$                                 \\ \hline
$\bm{m\ne 0}$ &  AI$^b$: NT [$U$: $\sT^{(2)}_S$]\phantom{HHH} &\phantom{HHH} AI: NT [$\sT_C$]\phantom{HHH} &\phantom{HHH} BDI: $\dZ$ [$\sC_C,\sT_C,\sX_C$] \\ \hline
$\bm{m=0}$ & BDI: $\dZ$ [$U$: $\sC^{(2)}_C\!,\sT^{(2)}_C\!,\sX^{(2)}_C$] & BDI: $\dZ$ [$\sC_{SC},\sT_C,\sX_S$] & BDI: $\dZ$ [$V$: $\sC^{(2)}_C\!,\sT^{(2)}_C\!,\sX^{(2)}_C$]            \\ \hline  \hline \rowcolor{Gray}[\tabcolsep]
$\rhomb$\textbf{-CSSH}             & $\bm{\phi=0}$                                   & $\bm{\phi\ne 0,\pi}$                       & $\bm{\phi=\pi}$                                 \\ \hline
$\bm{m\ne 0}$ & Trivial & AI: NT [$\sT'_{\subrhomb}$] & BDI: $\dZ$ [$\sC_C,\sT_{\subrhomb},\sX_{\subrhomb}$] \\ \hline
$\bm{m=0}$ & Trivial &  BDI: $\bm{\dZ}$ [$\sC'_{\subrhomb}$, $\sT'_{\subrhomb}$,$\sX_S$] & BDI: $\dZ$ [$W$: $\sC_W,\sT_W,\sX^{(2)}_S$]    \\ \hline \hline \rowcolor{Gray}[\tabcolsep]
$\Delta$\textbf{-CSSH }            & $\bm{\phi=0}$                                   & $\bm{\phi\ne 0,\pi}$                       & $\bm{\phi=\pi}$                                 \\ \hline
$\bm{m\ne 0}$ & Trivial & A: NT [none] & D: $\dZ_2$ [$\sC_C$] \\ \hline
$\bm{m=0}$ & Trivial & AIII: $\dZ$ [$\sX_S$] & BDI: $\dZ$ [$\sC_C,\sT_{SC},\sX_S$]
         \\ \hline
\end{tabular}

\caption{Symmetries and symmetry classes of the SSH, Creutz and CSSH Hamiltonians. Letters $\sC$, $\sT$ and $\sX$ refer to PH, TR and chiral symmetries, respectively. Hidden symmetries are marked by a tilde. Inversion symmetry is not included in the table, see text. All operators are defined in the text, and listed again in Appendix \ref{ap:syms}. In the CSSH models, $J\ne J'$ is assumed. When the Hamiltonian is block-diagonalizable using a unitary symmetry $U$, it is indicated by $U$:, followed by the symmetries of the resulting blocks. ``NT'' means the Altland-Zirnbauer classification predicts the model is not topological in 1D. Footnotes: $^a$ The hidden symmetries of the model change its symmetry class into a topological one and protect the edge states.
$^b$ The model only differs from the $m=0$ case in an element proportional to $\uno$, which has no effect on the topological invariant \cite{Perez-Gonzalez2019a} nor, in this case, the protected edge states.
}%
\label{tab:syms}
\end{table*}

\subsubsection{\texorpdfstring{$\phi\ne 0,\pi$}{phi!=0,pi}}

For other values of the magnetic flux, the TR symmetry holds but the PH and chiral symmetries break down. The edge states are only protected in the rungless case. When $m\ne 0$, the two edge states drift away from zero energy, given that the model belongs to class AI, with no protecting symmetries (see Table \ref{tab:syms}).
The Zak phase is still quantized due to inversion symmetry \cite{Asboth2015}, which also causes the end states to be degenerate in the thermodynamic limit. Thus, in this regime, the system is a topological crystalline insulator \cite{Li2015,Lau2018}. However, in it, the end states are not topologically protected against any kind of disorder, given that any disorder will break inversion symmetry. The topological phase diagram of the Creutz ladder as a function of $\phi$ and $m$ is depicted in Fig. \ref{fig:Zak} (a), as well as the regions where topological protection will be found in the CSSH ladders: whenever $\phi=\pi$ or $m=0$, like in the Creutz ladder, and, in the hourglass case, also when $\phi=0$.

When $m=0$, the topological protection can only be explained due to the presence of so-called ``hidden symmetries''. These are symmetries that satisfy one of the conditions (\ref{eq:sC}-\ref{eq:sX}), but they act in a non-trivial way on the spatial degree of freedom, so they cannot be reduced to their effect on the inner space Hamiltonian \cite{Cariglia2014}. They can determine the topology of the system, exactly like their conventional counterparts \cite{Hou2013,Fukui2013,Li2015B,Hou2016,Hou2017,Xiao2018,Marques2019}.

The topological protection when $m=0$ can be explained due to the presence of the hidden chiral symmetry $\tilde \sX_{S}$, which reflects the fact that the lattice is bipartite, and the hidden PH symmetry $\tilde \sC_{SC}$. The subindex $S$ indicates this symmetry is analogous to the chiral symmetry in the SSH chain, while $SC$ denotes that the operator is related to $\tilde \sX_S$ and $\sT^{(2)}_C$. In the spatial basis of the full Hilbert space, they are represented by the operators:
\begin{align}
    \tilde\sX_{S} = \diag\nolimits_{2L}(1,1,-1,-1,1,1,-1,-1,\ldots)\\
    \tilde \sC_{SC} =\diag\nolimits_{L}(\sigma_x,-\sigma_x,\sigma_x,-\sigma_x,\ldots)  \sK,
\end{align}
where $\diag_\ell (a,\ldots,z)$ is the diagonal or block-diagonal matrix formed by the $\ell$ elements $a,\ldots,z$. In general, we will mark operators acting on the whole Hamiltonian with a tilde, and those that act on the inner space Hamiltonian $\sH(k)$ with no tilde. These operators cannot be factorized to separate their effect on the inner and spatial subspaces, as is customary. 
These hidden symmetries place the rungless Creutz ladder in the BDI class (see Table \ref{tab:syms}).
The operators $\tilde\sX_S$ and $\sT^{(2)}_C$ are related to the hidden symmetries $\cS$ and $\sV$ mentioned in Ref. \cite{Li2015B} by the inversion symmetry of the ladder.

\subsubsection{AB caging and topological protection}

We want to bring attention to an important detail at this point, which is often overlooked. In most topological materials, if the system has a small size, the energies of the protected end modes have a finite (albeit small) value, due to their overlap. In the systems we study, these energies are at least one order of magnitude smaller than the gap, and so do not break the bulk-boundary correspondence. However, the zero modes in the rungless Creutz ladder are completely localized to the two end sites of the ladder, and this causes their overlap to vanish. This localization has the same nature as AB caging, as it is caused by destructive interference of the wavefunction due to the magnetic flux and connectivity of the ladder. Thus, the end modes are pinned at zero energy for any length of the ladder $L$, including the one plaquette limit, $L=2$. This interplay between AB localization and topology protects the zero modes even better than topology alone can, and allows the detection and use of protected states in systems with only a few sites. This also happens in the three CSSH ladders, where the edge states present a larger variety than in the Creutz case\footnote{The end states in the completely dimerized SSH chain also have zero energy for any length of the chain, as a trivial consequence of disconnecting the end sites from the rest of the chain. In the $\pi$-flux Creutz and CSSH ladders, this phenomenon occurs while the end sites are connected to the rest of the system.}.

\begin{figure}[htbp] 
    \includegraphics[width=0.95\linewidth,trim=10 10 10 10,clip]{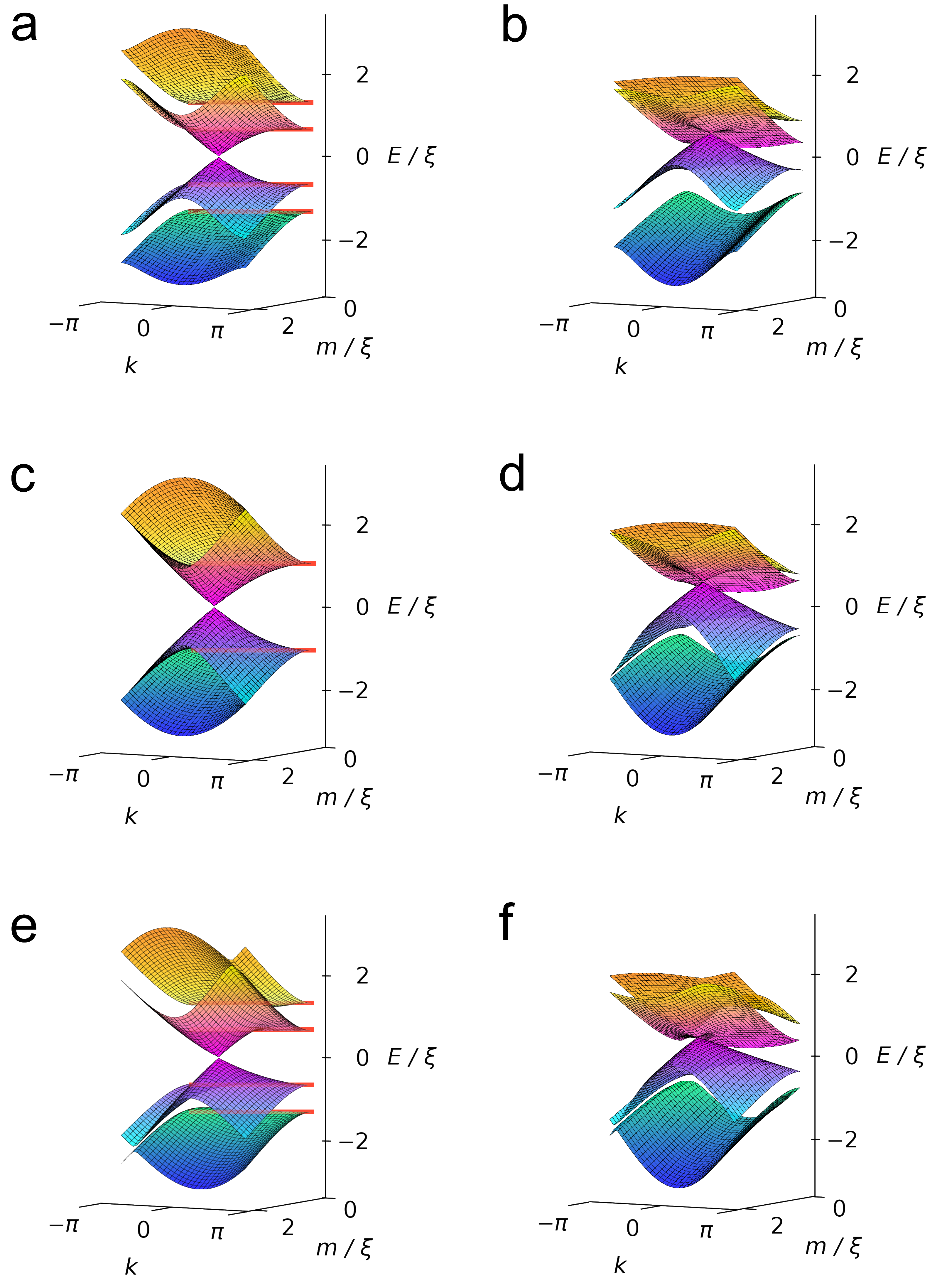}
    \caption{Bands of the CSSH ladders with periodic boundary conditions in $k$ space as a function of $m$. The parameters used are $\xi=1$, $J=2J'$, and $\phi=\pi$ for the figures on the left (a,c,e), and $\phi=\pi/2$ for those in the right (b,d,f). The former show flat bands when $m=0$ (marked in red), and a TPT when the middle bands touch at $k=0$. (a,b) $\hgls$-CSSH ladder spectra. The BDI case (a) has the most symmetric spectrum, while the AI case (b) is only symmetric in $k$, due to the TR symmetry. When $m=0$, the spectrum (b) is BDI. (c,d) $\rhomb$-CSSH ladder spectra. The BDI case (c) now has only one gap, with the two upper and the two lower bands touching at $k=\pi$ for all $m$, something that does not happen in the AI case (d). When $m=0$, the (d) case is also BDI. (e,f) $\Delta$-CSSH ladder spectra. Now, the $\pi$-flux spectrum (e) is only symmetric about the origin, indicating the sole presence of a PH symmetry (D class). For a generic value of $\phi$ (f), the spectrum shows no symmetries (A class), except for $m=0$, where only a chiral symmetry exists (AIII class), making the spectrum symmetric only in $E$.}
    \label{fig:3Dspectra}
\end{figure}

\subsubsection{\texorpdfstring{$\phi=0$}{phi=0}}

When $\phi=0 \mod 4\pi$, the Hamiltonian has a unitary commuting symmetry $U=\sigma_x$, that makes it decouple into two topologically trivial subspaces: a dispersive band spanned by $(\k{a}+\k{b})/\rdos$, and a flat band at energy $E=m$, spanned by $(\k{a}-\k{b})/\rdos$. When $\phi=2\pi \mod 4\pi$, the states that form each subspace switch their roles, and the value of their energies is reversed (for instance, the flat band is located at $E=-m$).

\section{Topology of the CSSH ladders} \label{sec:topo}

We will now analyze the three CSSH models. They have topological phases with protected zero modes if $\phi=\pi$ ($\pi$-flux regime) or $m=0$ (rungless regime). Additional constraints must be satisfied for each model to be topological, something we explore below using their topological phase diagrams (Fig. \ref{fig:Zak}). The coexistence of the symmetries from the $\pi$-flux and rungless regimes causes the $\phi=\pi, m=0$ case to be block-diagonalizable in the $\hgls$-CSSH and $\rhomb$-CSSH ladders, revealing a trivial and a topological subspace. The $\hgls$-CSSH ladder is also topological in the $\phi=0$ case, while the other two models are not. In the chiral models, two different nontrivial phases appear with edge modes of opposite chirality and winding number, in the same way than in the Creutz ladder \cite{Li2015,Sun2017}. These phases are separated by two topological phase transitions at $\phi=0, 2\pi \mod 4\pi$. For more details on the relevant topological invariants, please refer to Appendix \ref{ap:zak}.

\begin{figure*}[!htbp] 
    \includegraphics[width=\linewidth,trim=10 10 10 10,clip]{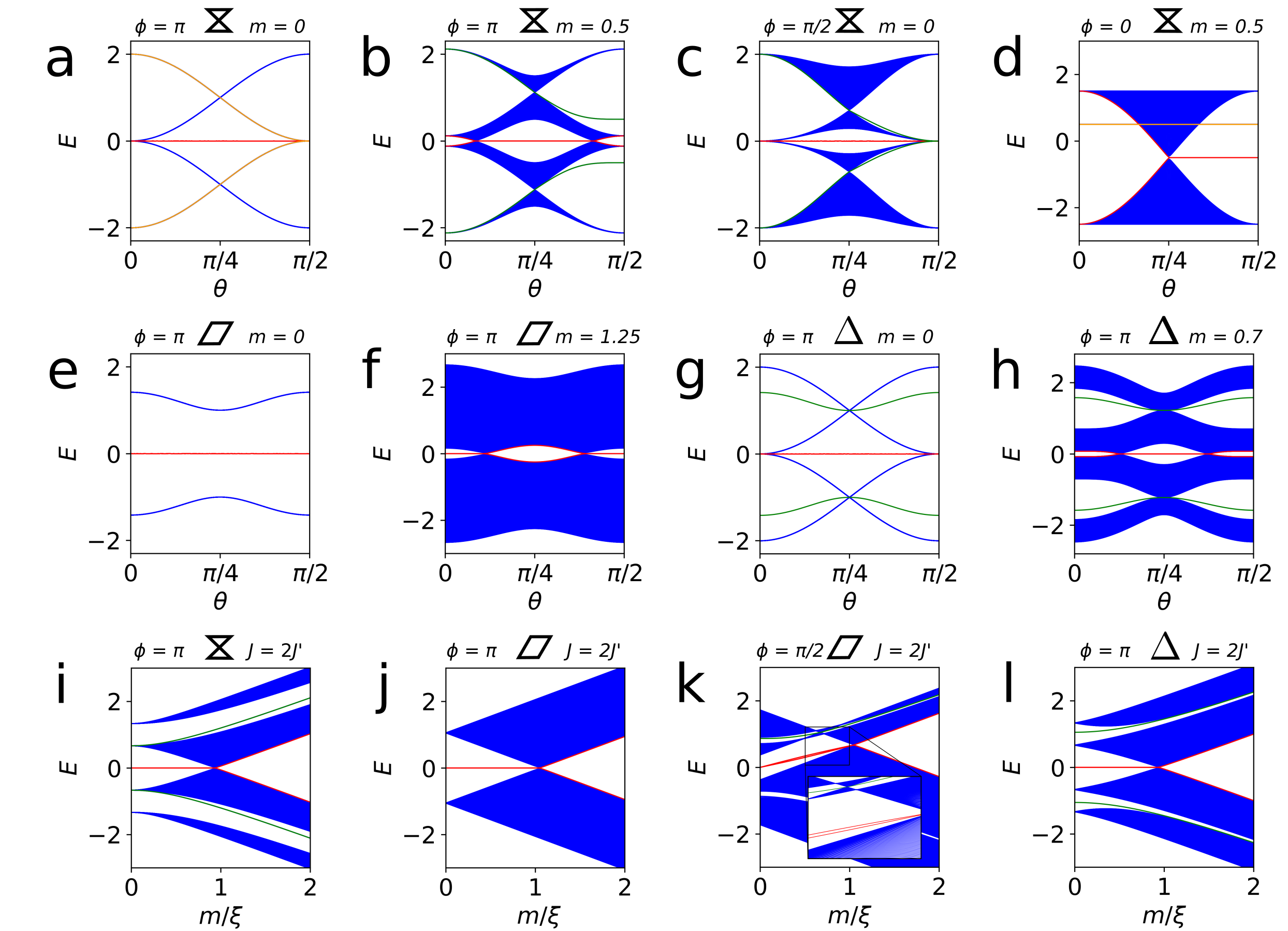}
    \caption{Spectra of different CSSH ladders with open boundary conditions as a function of $\theta$ (a-h) and $m$ (i-l). Bulk states are marked in blue or orange, protected zero modes in red, and other edge states in green. All systems have $L=200$ and $\xi=1$, and were obtained by exact diagonalization. The parameters and CSSH family of each subfigure are included above it. (a-d) $\hgls$-CSSH ladder spectra in $\theta$. (a) Flat-band regime. The band energies can be tuned independently changing $J$ and $J'$, and they correspond to a trivial (orange) and a topological (blue) completely dimerized SSH system (see text). (b) General BDI $\pi$-flux case. Zero modes and non-protected topological edge states appear. Two TPTs occur at the points where the zero modes join the bulk bands.  (c) BDI $\pi/2$-flux rungless topological phase, with protected zero modes. (d) Fluxless BDI case, showing the SSH-like subspace with a TPT at $\theta=\pi/4$, and a degenerate flat band (orange). (e,f) $\rhomb$-CSSH ladder spectra in $\theta$. (e) Flat-band case, always topological. (f) BDI $\pi$-flux case. The two upper and lower bands touch, so only one gap appears. When it closes, there is a TPT. (g,h) $\Delta$-CSSH ladder spectra in $\theta$. (g) Flat-band regime. Non-protected end states appear in addition to the zero modes. (h) D-class $\pi$-flux case, with non-protected end states and two TPTs. (i-l) Spectra as a function of $m$. (i) $\pi$-flux $\hgls$-CSSH spectrum. A TPT can be seen at $m=1$, and non-protected edge states come out of the inner bands as $m$ increases. (j) $\pi$-flux $\rhomb$-CSSH ladder spectrum. A TPT occurs at the same point as in (i). The two upper and lower bands touch, so only one gap is seen. (k) $\rhomb$-CSSH ladder spectrum for $\phi=\pi/2$. When $m\ne 0$, the edge states in the central gap are not degenerate, because they are not related by inversion symmetry. The color red has been used for clarity. The third gap undergoes a phase transition (see insert). (l) $\pi$-flux $\Delta$-CSSH ladder spectrum. A TPT is also found at $m=1$, and non-protected end states go into the outer bands as $m$ increases. In (i,j,l), the flat-band regime can be seen at $m=0$.}
    \label{fig:2DspectraU}
\end{figure*}

The different topological classes, symmetries and invariants of the SSH, Creutz and CSSH systems for every combination of parameters can be found in Table \ref{tab:syms}. The detailed list with the analytical expression of all symmetry operators is reserved for Appendix \ref{ap:syms}, although the most relevant ones will be mentioned in the text. The bulk bands for the different models, depicted in Fig. \ref{fig:3Dspectra}, reflect the presence of these symmetries. The spectra for the finite ladders, with different types of topological end modes, is also studied below (Fig. \ref{fig:2DspectraU}). Finally, the spatial distribution of some energy eigenstates of different 32-site ladders are represented in Fig. \ref{fig:states}. Each of the aforementioned figures is referenced below, in the context of the different models. In all cases, the energy scale has been fixed by setting $\xi=1$.

\subsection{The hourglass CSSH ladder}

\subsubsection{\texorpdfstring{$\phi=\pi$}{phi=pi}}
In this case, the $\hgls$-CSSH ladder [see Fig. \ref{fig:CSSH} (c)] belongs to class BDI, with symmetries $\sC_C=\uno_2 \otimes \sigma_z \sK$, $\sT_C=\uno_2\otimes \sigma_x \sK$ and $\sX_C=\uno_2\otimes (i\sigma_y)$, which are equivalent to the Creutz ladder symmetries in the same regime, and have dimension 4 (the superindex $(4)$ was omitted in all matrices for clarity). This is reflected on the symmetric form of its spectrum [see Fig. \ref{fig:3Dspectra} (a)].

In the rungless $\pi$-flux case, the system has four flat bands and a commuting unitary symmetry $V=\sigma_z\otimes (i\sigma_y)$. This kind of symmetry allows us to block-diagonalize the Hamiltonian, and study the topology of the different subspaces independently. In this case, the system can be mapped into two independent, fully dimerized SSH chains, a trivial and a topological one, with amplitudes $v_1=2J'$, $w_1=0$ and $v_2=0$, $w_2=2J$. They correspond, respectively, to the flat bands with energies $E=\pm 2J'$ and $E=\pm 2J$ [see Fig. \ref{fig:2DspectraU} (a)]. The $\hgls$-CSSH ladder end modes correspond to the edge states of the topological chain. When $m$ is switched on, it couples the two chains and the symmetry $V$ breaks.

In the $m\ne 0$ case, the full four-band system must be considered. The topological invariant related to the boundary states that appear in a particular gap in the system can be obtained as the sum of the Zak phases of all the bands below it \cite{Wang2018}, $\tilde \sZ_i = \sum_{j=1}^i \sZ_j$. In a topological symmetry class, the number of topological edge states per end in that gap will be $\tilde \nu_i = \tilde \sZ_i/\pi$ \cite{Wang2019,Wang2018,Zhu2018,Deng2013}. When quantized, the topological invariants associated with the first and third gaps of all CSSH ladders have the same value, $\tilde\sZ_1=\tilde\sZ_3$, so to identify different phases we will use $[\tilde \sZ_1,\tilde \sZ_2]$.

The topological phase diagram for this regime is depicted in Fig. \ref{fig:Zak} (b). The first invariant, $\tilde \sZ_1$, behaves like the one in the SSH chain, being topological when $J>J'$. The second one, $\tilde \sZ_2$, is topological when $m < \xi \sin(2\theta)=2\sqrt{JJ'}$. When $J=J'$, this condition is reduced to the one in the Creutz ladder. The states in the second (or central) gap are protected by the chiral and PH symmetries in the topological phase [see Fig. \ref{fig:2DspectraU} (i)]. Those in the first and third gaps only appear when $\tilde\sZ_1=\pi$, and thus satisfy the bulk-boundary correspondence. However, they are not pinned in the middle of their gaps, as can be seen in Fig. \ref{fig:2DspectraU} (b). This is because the symmetries of the system can only protect states in the middle gap, given that they only guarantee a symmetric spectrum about zero energy. This is further explained in Appendix \ref{ap:prot}. They do guarantee, however, that the states in the first gap have the opposite energy to those in the third gap. These states can be described as topological, but not protected. If $\pi/4<\theta < \arcsin(m/\xi)/2$, the system has six topological edge states, two of which are protected at zero energy. In Fig. \ref{fig:states} (b,c), the edge states in the first and second gap of the $\phi=\pi, \theta=3\pi/8, m=0.5$ system are shown.

\subsubsection{\texorpdfstring{$\phi\ne 0,\pi$}{phi!=0,pi}}
For these values of the magnetic field, the TR symmetry still holds, but the other two break down, in a similar situation to that of the Creutz ladder. This can be appreciated in Fig. \ref{fig:3Dspectra} (b). If $m \ne 0$, the model is a topological crystalline insulator in class AI. Its states are not topologically protected against disorder, but they are still degenerate in a long enough ladder due to spatial inversion.

When $m=0$, the same chiral and PH symmetries as in the Creutz ladder appear. However, they are not hidden anymore, because the unit cell of the system is larger now. Their operators on the bulk Hamiltonian are $\sC_{SC}=\sigma_z\otimes \sigma_x \sK$ and $\sX_S=\sigma_z\otimes \uno_2$. As noted earlier, the subindex $S$ indicates this symmetry is analogous to the chiral symmetry in the SSH chain. In this regime, $\tilde\sZ_1=\pi$ only when $J>J'$, and $\tilde\sZ_2=\pi$ always holds. Like in the $\phi=\pi$ case, only the states in the middle gap are protected [see Fig. \ref{fig:2DspectraU} (c)]. Their exact form is described in Appendix \ref{ap:edge}. The Zak phases take the same values as in the $\pi$-flux regime [Fig. \ref{fig:Zak} (b)]. 

\subsubsection{\texorpdfstring{$\phi=0$}{phi=0}}
If the magnetic field is turned off in the $\hgls$-CSSH ladder, the topology inherited from the Creutz ladder disappears, but not the one generated by the staggered, SSH-like amplitudes. In this case, the bulk Hamiltonian has a unitary commuting symmetry, $U=\uno_2\otimes \sigma_x$, that generates two independent subspaces, exactly like in the Creutz case. One of them is a topologically trivial doubly degenerate flat band at energy $E=m$, spanned by the internal states $(\k{A}-\k{B})/\rdos$ and $(\k{C}-\k{D})/\rdos$. This subspace can be turned into two non-degenerate flat bands if the vertical hopping amplitudes $m_1,m_2$ are set to different values in \eqref{eq:CSSH}.

The other subspace is now analogous to an SSH chain with amplitudes $v=J',w=J$, which will be topological if $J'<J$, in contrast to the Creutz case, where the system is metallic at $\phi=0$. Its spectrum is pictured in Fig. \ref{fig:2DspectraU} (d). The vertical amplitude $m$ behaves like a homogeneous chemical potential, shifting the SSH chain spectrum downwards, but having no other effect in the topology. The topological chiral end states are analogous to those in the SSH chain, with the left (right) state only supported on the odd (even) sites, as can be seen in Fig. \ref{fig:states} (d) for the $\theta=3\pi/8$, $m=0$ system.

\subsection{The rhomboid CSSH ladder}

\subsubsection{\texorpdfstring{$\phi=\pi$}{phi=pi}} \label{sec:Rhomb}
When $\phi=\pi$, the $\rhomb$-CSSH ladder [see Fig. \ref{fig:CSSH} (d)] is in class BDI, with the symmetries
\begin{align}
    &\sC_C=\uno_2\otimes \sigma_z \sK\\
    &\sT_{\subrhomb} = \begin{pmatrix}
    -is_\varth & c_\varth & 0 & 0\\
    c_\varth & -i s_\varth & 0 & 0\\
    0 & 0 & 0 & 1\\
    0 & 0 & 1 & 0
    \end{pmatrix} \sK\\
    &\sX_{\subrhomb} = \begin{pmatrix}
    is_\varth & c_\varth & 0 & 0\\
    -c_\varth & -i s_\varth & 0 & 0\\
    0 & 0 & 0 & 1\\
    0 & 0 & -1 & 0
    \end{pmatrix}
\end{align}
where
\begin{align}
    s_\varth = \sin \varth = \frac{J'^2-J^2}{J'^2+J^2}\\
    c_\varth = \cos \varth = \frac{2J'J}{J'^2+J^2}.
\end{align}
The angles $\theta$ and $\varth$ are related by the expression:
\begin{equation}
    \varth = 2\arctan(\cos 2\theta).
\end{equation}

The first and third gaps are closed at $k=\pi$, forming two Dirac cones [see Fig. \ref{fig:3Dspectra} (c)], and this renders $\tilde\sZ_1$ and $\tilde\sZ_3$ not meaningful, as only the middle gap exists. The ground state of the system lies at $k=0$, something apparent in Fig. \ref{fig:states} (e) for $J=2J'$, $m=0.5$.
The relevant phase diagram is shown in Fig. \ref{fig:Zak} (c). The system is topological when $m<\sqrt{2(J^2 + J'^2)}=\xi\sqrt{(3+\cos 4\theta)/2}$. When that condition is met, the gap has zero modes, protected by the chiral and PH symmetries. Their form for $J=2J'$, $m=0.5$ is depicted in Fig. \ref{fig:states} (f), and they can be seen in its spectra as a function of $\theta$, in Fig. \ref{fig:2DspectraU} (f), and as a function of $m$, in Fig. \ref{fig:2DspectraU} (j).

The chiral symmetry of the system $\sX_{\subrhomb}$, which protects the end modes along with $\sC_C$, depends on $\theta$. It interpolates between a Creutz-like chiral symmetry, at $\theta = \pi/4$, and its counterpart in the orthogonal dimer chain, at $\theta = 0,\pi/2$. This provides a degree of freedom to choose the system in which the protecting symmetries will be less broken in a particular implementation. The Creutz ladder chiral symmetry can withstand a broader range of on-site disorder configurations without breaking than that in the orthogonal dimer chain (see Appendix \ref{ap:syms} for a more detailed discussion). On the other hand, the orthogonal dimer chain is a simpler model, and can therefore have less sources of disorder than the Creutz ladder, so it might be preferable for some applications.

As can be seen in Fig. \ref{fig:Zak} (c), when $J$ or $J'$ vanish, the system can still be topological, unlike the other models. The resulting lattice is the orthogonal dimer chain, depicted in Fig. \ref{fig:CSSH} (f). It is topological when $\phi=\pi$ and $m<\sqrt{2} J$ (assuming $J'=0$). As mentioned above, this model is usually studied in the context of interacting spins \cite{Richter1998, Paulinelli2013, Verkholyak2013, Nandy2019, Galisova2021}. The $\rhomb$-CSSH ladder connects the topological phases of the Creutz ladder (obtained when $\theta=\pi/4$) and the orthogonal dimer chain ($\theta=0,\pi/2$), two previously unrelated models. The superposition of both topological edge states in the orthogonal dimer chain with $m=0.5$ is shown in Fig. \ref{fig:states} (i).

Furthermore, when $\phi=\pi$ and $m=0$ (for any value of $\theta$), the system has a unitary commuting symmetry $W=\sX_S \sX_{\subrhomb}$ that creates two subspaces that can be mapped to two fully dimerized SSH chains, with amplitudes $v_1=0=w_2, w_1 = i\sqrt{2(J^2+J'^2)}=-v_2$. They are topological and trivial, respectively. The complex amplitudes, corresponding to a magnetic potential in the direction of the chain, do not affect the topology, and only modify the form of the PH and TR operators of the SSH chain. The topological subspace has two protected zero modes. The upper $(+)$ and lower $(-)$ bands of the two SSH chains are degenerate, with energies $E=\pm\sqrt{2(J^2+J'^2)}$. However, given that their states never mix, there is no need to resort to the non-abelian topological invariant \cite{Stanescu2016}. The spectrum of the system is pictured in Fig. \ref{fig:2DspectraU} (e).

\subsubsection{\texorpdfstring{$\phi\ne 0,\pi$}{phi!=0,pi}}
The system is still TR-symmetric for these values of $\phi$ [see Fig. \ref{fig:3Dspectra} (d)], but the form of the TR operator is now more complicated (see equation \eqref{eq:TTun} in Appendix \ref{ap:syms}). When $m=0$, the chiral symmetry $\sX_S$ pins the edge states in the middle gap exactly at zero energy. The left zero mode is depicted in Fig. \ref{fig:states} (g) for $J=2J'$, $\phi=\pi/2$. A PH symmetry, $\sC'_{\subrhomb}$ (see Appendix \ref{ap:syms}), is also present there. Therefore, the $m=0$ case belongs to the BDI class, while the $m\ne 0$ system belongs to class AI, which is not topological.

To understand the topological phase diagram, we need to keep in mind that the Zak phase is only guaranteed to be quantized if the system with periodic boundary conditions is symmetric under a spatial inversion operation that leaves the point at the center of the unit cell invariant \cite{Marques2019}. The $\rhomb$-CSSH ladder is symmetric under a specific spatial inversion symmetry, in which the inversion center is located between the $C$ and $D$ sites (see Appendix \ref{ap:syms} for a matrix representation). 

In this regime, the invariants associated with the first and third gaps are no longer equal. In fact, one of them has a topological phase transition between non-quantized values, while the other one does not. Only $\tilde\sZ_3$ presents a phase transition if $-\pi<\phi<\pi \mod 4\pi$, and only $\tilde\sZ_1$ presents one otherwise. This can be seen in Fig. \ref{fig:Zak} (d). Non-protected end states appear in the first and third gaps, and they do not follow the usual bulk-boundary correspondence. One of these end states is pictured in Fig. \ref{fig:states} (h).
The topological phase transitions are related to gap closings in the spectrum that do not change the number of end states [see the insert in \ref{fig:2DspectraU} (k)], but do change the end (left or right) in which they appear. We discuss these states in depth in Section \ref{sec:DStates}.

In the $\rhomb$-CSSH ladder, the Zak phase is quantized whenever the system belongs to a topological class, except for $\tilde \sZ_1$ and $\tilde \sZ_3$ in the $\phi\ne\pi,m=0$ case. There, $\tilde \sZ_2$ is equal to $\pi$, indicating the presence of topological zero modes.  
Additionally, the second gap end states in a ladder with an even number of rungs are not related by spatial inversion, and so they are not degenerate, even in a large system. This occurs because the open system loses its global inversion symmetry, and its effects can be seen in the corresponding spectrum as a function of $m$ in Fig. \ref{fig:2DspectraU} (k).

\subsubsection{Edge state tunability}

\begin{figure*}[!htbp] 
    \includegraphics[width=\linewidth,trim=10 10 10 10,clip]{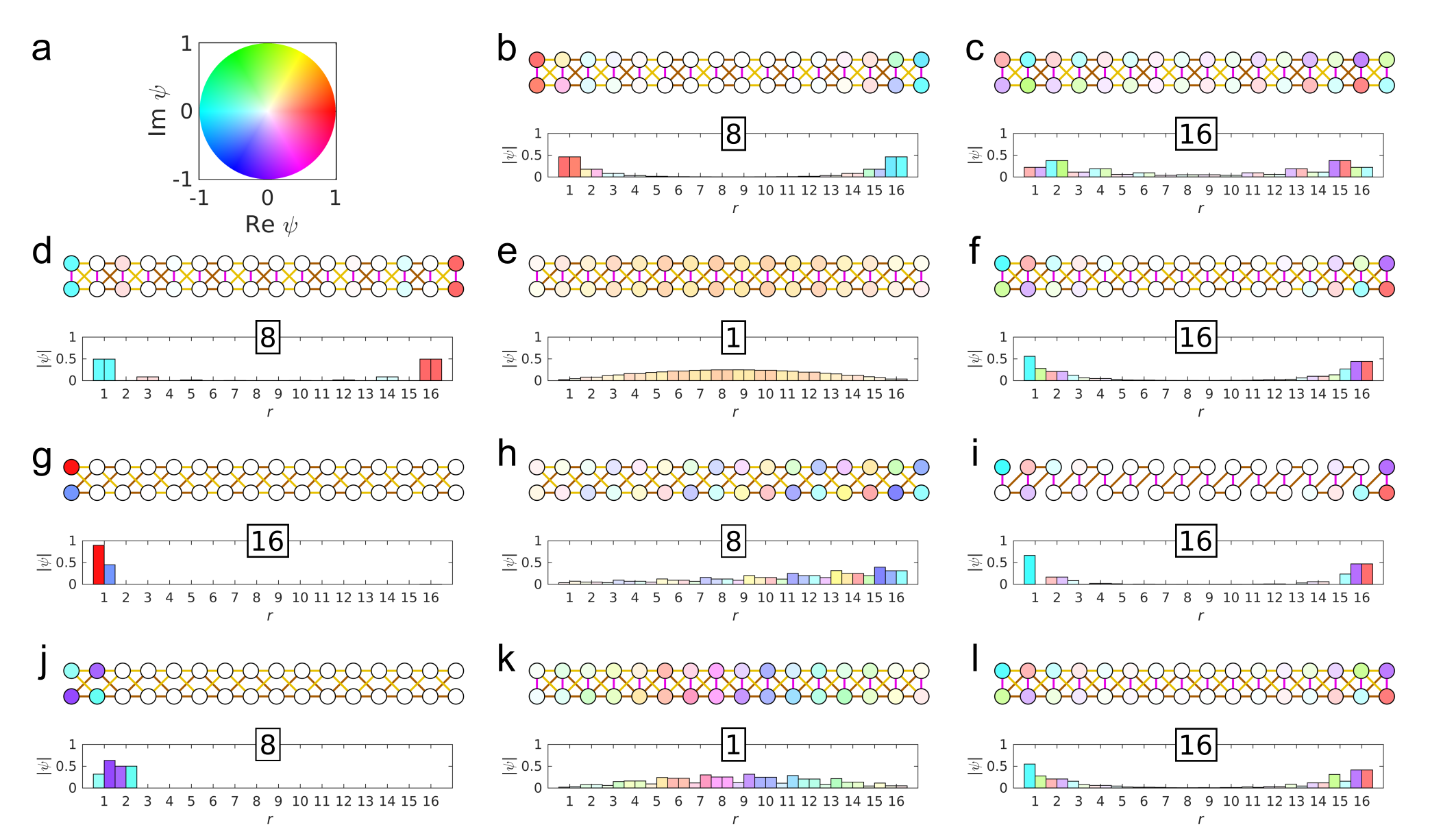}
    \caption{Spatial distribution of the wavefunctions of different energy eigenstates in different 32-site ladders with open boundary conditions, with $\xi=1$. We identify the ladders by their symbol and parameters $(\theta,\phi,m)$, and the states by its number in a square, by ascending order of energy.
    (a) Color code used to show the phase of the wavefunction. (b,c) $\hgls(3\pi/8,\pi,0.5)$ ladder, edge states in the first gap (b) and in the second gap (c). 
    (d) $\hgls(3\pi/8,0,0.5)$ ladder, protected zero mode. Its profile is reminiscent of an SSH chain edge state. 
    (e,f) $\rhomb(0.96,\pi,0.5)$ ladder. The value of $\theta\approx 0.96$ was chosen such that $J=2J'$. Its ground state (e) lies at $k=0$ [see Fig. \ref{fig:3Dspectra} (c)], and its topological zero modes (f) are similar to those in the Creutz ladder, but have a larger support on the upper leg.
    (g,h) $\rhomb(0.96,\pi/2,0)$ ladder left AB-caged zero mode (g) and right non-protected state (h). The left zero mode (g) can be tuned using $\theta$ and $\phi$ to take almost any form in the Bloch sphere, and is degenerate with its left counterpart (see text). The relative phase between its sites is $\phi/2=\pi/4$. 
    (i) $\rhomb(\pi/2,\pi,0.5)$ ladder topological state. This model has $J'=0$ and is identical to the orthogonal dimer chain. 
    (j) $\Delta(0.96,\pi,0)$ ladder AB-caged non-protected state in the first gap. It is localized in the four left sites due to AB caging.
    (k,l) $\Delta(0.96,\pi,0.5)$ ladder (D-class TI). The loss of chiral and TR symmetries now shifts the ground state of the system (k) to a nonzero value of $k$ [see Fig. \ref{fig:3Dspectra} (e)]. This is reflected on the changing phase of the wavefunction as a function of $r$ [compare to (e)]. The states in the central gap (l) are now only protected by PH symmetry.}
    \label{fig:states}
\end{figure*}

To show an application of the high tunability of the protected edge states, let us consider the left topological zero mode of the rungless $\rhomb$-CSSH ladder [see Fig. \ref{fig:states} (g)]:
\begin{equation} \label{eq:Lrh}
    \k{\sL_{\subrhomb}} \!=\! \frac{J\k{1,A}-J' e^{i\phi/2}\k{1,B}}{\sqrt{J^2 + J'^2}}.
\end{equation}
The dependence on $\phi$ and $\theta$ of the zero modes provides a topologically protected method to control a particle located in the two end sites, by changing the parameters of the system.

We can establish any state in the Bloch sphere, with $\theta$ controlling the weight of each site and $\phi$ controlling their relative phase. To obtain a given state $\k{\psi}=a\k{1,A}+b e^{i\varphi}\k{1,B}$ with $a,b,\varphi\in\dR$, the following parameters must be chosen:
\begin{align}
    &\phi_\psi = 2(\varphi-\pi) \mod 4\pi\\
    &\theta_\psi = \arctan \sqrt{\frac{a}{b}},
\end{align}
such that $a=J/\sqrt{J^2+J'^2}$, $b=J'/\sqrt{J^2+J'^2}$. In the cases of $\varphi=0,\pi$ ($\phi=0 \mod 2\pi$), the system becomes topologically trivial. The states defined by expression \eqref{eq:Lrh}, however, are still eigenvalues of the model, and become part of a flat band instead of being topological zero modes.

In particular, the system can be initiated with the particle in eigenstate $\k{1,A}$ in the model with $J'=0$ and $\phi=\phi_\psi$, and then turning $J'$ on adiabatically until the condition $\theta=\theta_\psi$ is reached. In particular, this allows us to prepare the topological end modes of the Creutz ladder, by choosing $\theta = \pi/4$. As long as the chiral symmetry of the system, $\sX_S$, is not broken, this protocol is topologically protected against disorder, even in small systems. The left state in the $\Delta$-CSSH ladder can also be used, with the same results. For more information, refer to Appendix \ref{ap:edge}, where we discuss the caged zero modes of all models.

\subsubsection{\texorpdfstring{$\phi=0$}{phi=0}}

In this regime, the system is topologically trivial. When $m=0$, a doubly degenerate flat band appears at $E=0$, and one dispersive band is located at each side of it, touching it if $\theta=\pi/4$. When $m$ is turned on, a flat band remains at $E=m$, spanned by $(\k{C}-\k{D})/\rdos$, and the other three are dispersive. 

\subsection{The delta CSSH ladder}
\subsubsection{\texorpdfstring{$\phi=\pi$}{phi=pi}}

When $m\ne 0$, this $\pi$-flux model, in contrast with all the previous ones, does not have TR and chiral symmetries inherited from the $\phi=\pi$ Creutz ladder, and only has its PH symmetry, $\sC_C$. This is related to the fact that one of the legs is formed by strong links, and the other one by weak ones [see Fig. \ref{fig:CSSH} (e)]. Given that the TR symmetry in Creutz-related ladders swaps the two legs, this imbalance breaks TR symmetry. The effects of this symmetry breaking can be easily seen in its spectrum [Fig. \ref{fig:3Dspectra} (e)], which is symmetric about the origin $k=0,E=0$, but not in $E$ or in $k$. The first and third gaps are narrowest at opposite values of $k$, and not in $k=\pi$ as in the other ladders.

Therefore, the $\Delta$-CSSH ladder with $\phi=\pi$ and $m\ne 0$ belongs to the D class. Other instances of 1D or quasi-1D topological D-class materials are typically superconducting systems. 
Its topology is characterized by a $\sZ_2$ invariant, which in the 1D D class coincides with the Zak phase\footnote{Its value for the middle gap can also be calculated using the Pfaffian of the Hamiltonian. This has been confirmed numerically to coincide with $\tilde \sZ_2$ in the $\Delta$-CSSH ladder.} \cite{Budich2013,Li2016}.
As depicted in Fig. \ref{fig:Zak} (e), $\tilde\sZ_1$ takes a constant value of $3\pi/2$, except in the Creutz limit $\theta=\pi/4$, where it becomes undefined. On the other hand, $\tilde\sZ_2$ indicates that protected zero modes will appear if $m<2\sqrt{JJ'}$. These states are represented in Fig. \ref{fig:states} (k,l) for $J=2J'$ and $m=0.5$. They are only protected by PH symmetry, as there is no chiral symmetry in the system.

Additionally, non-protected edge states located in the left end of the ladder appear in the first and third gaps. They can be seen in the spectrum in Fig. \ref{fig:2DspectraU} (g,h,l). One state of this kind in the flat band case is depicted in Fig. \ref{fig:states} (j). Interestingly, these states do not follow the usual bulk-boundary correspondence, and can be related instead to a Chern number. We discuss these states in detail in subsection \ref{sec:DStates}, and in Appendix \ref{ap:chern}.


As usual, when $\phi=\pi$ and $m=0$, the bands flatten and the chiral and TR symmetries $\sX_S$ and $\sT_{SC}$ appear (see Appendix \ref{ap:syms}). The model is BDI and topological ($\tilde\sZ_2=\pi$). The end states of the system are now totally localized to two sites due to AB caging.

\subsubsection{\texorpdfstring{$\phi\ne 0,\pi$}{phi!=0,pi}}
In the $m\ne 0$ case, the model has no relevant symmetries [see Fig. \ref{fig:3Dspectra} (f)] and belongs to class A, not topologically protected in 1D. The second gap end modes are not degenerate even in the even-runged, large chain case, because the boundaries of the system break spatial inversion. In this case, $\tilde\sZ_1 \approx \tilde\sZ_3$, so we only represent $[\tilde\sZ_1,\tilde\sZ_2]$ in the phase diagram [Fig. \ref{fig:Zak} (f)].

When $m=0$, the chiral symmetry $\sX_S$ puts the system in class AIII and makes the system topological, with $\tilde\sZ_2=\pi$ and $\tilde\sZ_1 = 3\pi/2$, and hosts a pair of protected zero modes in the middle gap (see Appendix \ref{ap:edge}). Like their $\rhomb$-CSSH counterparts, these states are localized in only two sites, and they can be tuned in the same way as them. Non-protected states of the same nature as in the $\pi$-flux can also appear, see the next subsection for details. 
There are other known ways to achieve AIII TIs with the geometry of the Creutz ladder: to include an energy imbalance between the legs \cite{Piga2017}, or additional complex hopping amplitudes in the diagonal or vertical tunneling amplitudes \cite{Velasco2019,Sun2017,Kuno2020}. As we have seen, by changing its parameters, the $\Delta$-CSSH ladder can belong to three of the five topological classes in 1D (see Table \ref{tab:clas}). The end modes remain protected when changing from one class to another.

The Zak phase is not always quantized in the $\Delta$-CSSH ladder because, like the $\rhomb$-CSSH case, it is only invariant under a specific kind of generalized inversion symmetry. This transformation only flips the longitudinal coordinate of the ladder (see Appendix \ref{ap:syms}). The value of $\tilde\sZ_2$ is quantized when the system belongs to a topological symmetry class, but not $\tilde\sZ_1$, as we discuss in Section \ref{sec:DStates}.

\subsubsection{\texorpdfstring{$\phi=0$}{phi=0}}
The system is topologically trivial. 
Like in the other models, a doubly degenerate flat band appears in $E=0$ in the rungless regime. The other two bands are dispersive and touch the flat band for all values of $\theta$, making the system metallic. Like in the $\rhomb$-CSSH case, when $m$ is switched on, a non-degenerate flat band spanned by $(\k{C}-\k{D})/\rdos$ lies at $E=m$, while the rest of the bands become dispersive.

\subsection{Non-protected states and Chern numbers}\label{sec:DStates}
In the rhomboid and delta CSSH ladders, a single edge state can be found in the first or in the third gaps, and they often appear in both at the same time. These states, when present, are both localized at the same side of the ladder, not one at each side. They can be found at the left or right ends in the $\rhomb$-CSSH ladder, and only at the left end in the $\Delta$-CSSH ladder. These states are not protected, as they are not located in the central gap (see Appendix \ref{ap:prot} for details), and they do not follow the usual bulk-boundary correspondence, as they appear even though $\tilde\sZ_1$ takes intermediate values between $0$ and $\pi$ [see Fig. \ref{fig:Zak} (d,e,f)]. However, they have a topological nature, given that they appear due to a topological obstruction that can be understood by exploring the Wannier functions of the model. For a detailed analysis, please refer to Appendix \ref{ap:chern}.

Interestingly, this kind of edge states have been recently reported in a recent paper \cite{Martinez2019} in trimer chain models with no inversion symmetry. In said work, it is argued that these edge states, even though they exist in a 1D model, are related to the 2D Chern number associated to the model changing along a closed path in its space of parameters. This is associated with the topological pumping that would happen if the model was physically changed in time following the chosen path \cite{Martinez2019,Ke2016}.

\begin{figure}[!htbp] 
    \includegraphics[width=\linewidth,trim=10 10 30 5,clip]{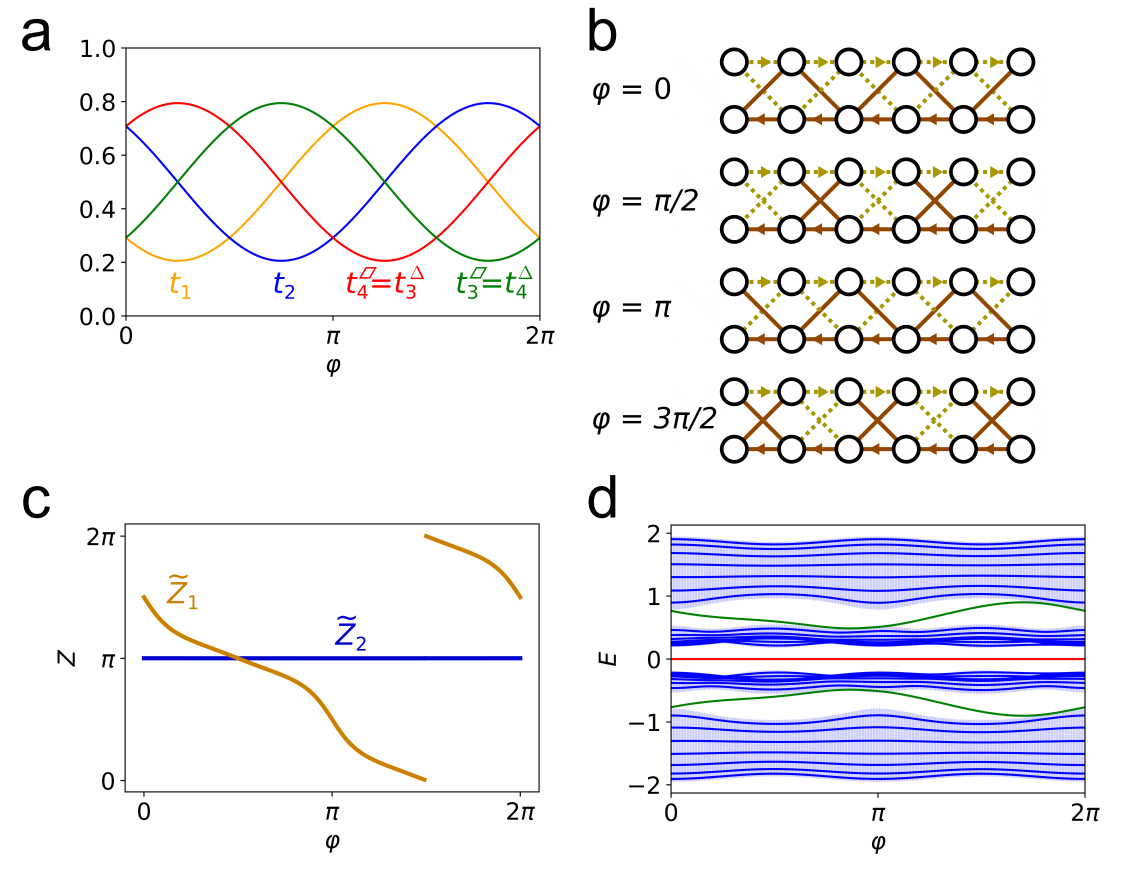}
    \caption{(a) Hopping parameters along the path for the calculation of the Chern number, for $J=0.7$ and $J'=0.3$. (b) Scheme of the hopping parameters for the delta case as a function of $\vf$, where the darker links have an amplitude of $J$ and the lighter, dotted lines take the value $J'$. (c) Zak phase of the first and second gaps along the path in the $\theta=1,\phi=\pi/2,m=0$ $\Delta$-CSSH ladder. $\tilde\sZ_1$ winds around once in the negative direction, and so $C_1=-1$, while $\tilde\sZ_2$ does not ($\tilde C_2 = 0$). (d) Spectrum with periodic (light blue, in the background) and open boundary conditions (lines) for the system in (c). The end states in the first and third gaps (green) move around as dictated by its Chern number, while those in the central gap (red) remain pinned at zero energy. The finite system has $L=16$.}
    \label{fig:Chern}
\end{figure}

In the CSSH ladders, an analogous invariant can be found, if we consider an appropriate extension of the models. In the $\rhomb$-CSSH ladder, the relevant path along parameter space can be chosen as [see Fig. \ref{fig:Chern} (a)]:
\begin{align}
    &t_1 = A_0 + A\sin (-\frac{3\pi}{4} +\vf) \nonumber\\
    &t_2 = A_0 + A\sin (\frac{3\pi}{4} +\vf) \nonumber\\
    &t_3 = A_0 + A\sin (-\frac{\pi}{4} +\vf) \nonumber\\
    &t_4 = A_0 + A\sin (\frac{\pi}{4} +\vf)
\end{align}
where $A_0 =(J+J')/2$, $A=(J-J')/\sqrt{2}$, $0<\vf<2\pi$ is the parameter along the path, $t_i$ are the diagonal hopping amplitudes in Hamiltonian \eqref{eq:CSSH}. The rest of the hopping amplitudes are constant along the path. For the $\Delta$-CSSH ladder, the value of $t_3$ and $t_4$ must be interchanged.

For $\vf=0$, the system is a $\rhomb$- or $\Delta$-CSSH ladder as defined in Section \ref{sec:CSSH}. In the opposite point of the path, $\vf=\pi$, the system is the result of reflecting said ladder along the horizontal axis, and thus the value of its Zak phase will be the opposite. This system will have non-protected end modes on the opposite side to the non-reflected version. As an example, the different models visited along the path in the delta case are depicted in Fig. \ref{fig:Chern} (b).

For the $\phi=\pi$ $\Delta$-CSSH ladders, the PH symmetry must be broken at some points to avoid band degeneracies along the path, and so in that case we also consider a changing $\phi$, for instance:
\begin{equation}
    \phi(\vf) = \frac{\pi}{4} [3+\cos(2\vf)].
\end{equation}
Thus, $\phi(\vf=0,\pi) = \pi$, and the degeneracies that otherwise appear at $\vf=\pi/2,3\pi/2$ are avoided. In the rhomboid case, the first and third gaps of the $\pi$-flux systems close, so the invariant loses meaning there.

As the system changes along the path, the Zak phase of the first band starts increasing or decreasing, and eventually wraps around and returns to its initial value. This winding of the Zak phase indicates that the lowest band has a nontrivial Chern number as it changes along the path. In the $\Delta$-CSSH ladder, $C_1 = \pm 1$ when $\theta \lessgtr\pi/4$. We can see this for $\theta=1,\phi=\pi/2,m=0$ in Fig. \ref{fig:Chern} (c), where $C_1=-1$.
The second gap has an invariant of $\tilde C_2 = C_1 + C_2 = 0$, which predicts no edge states of this kind there (the states in this gap are related to the usual Zak phase instead). Equivalently, the value of $\tilde\sZ_2$ along the path does not wrap around, and instead remains constant [see Fig. \ref{fig:Chern} (c)]. The value of $\tilde C_3 = C_1 + C_2 + C_3$ is nontrivial, indicating the possible existence of an edge state in the third gap.


It is important to note that invariant is associated with the whole path, not just the particular system we are interested in. For this reason, the bulk-boundary correspondence is only guaranteed to be satisfied when considering the entire path, not at a single point in it. The value of the Chern number cannot predict if the end states in the CSSH ladders will appear in the gaps or not, it only indicates that a single state \textit{can} appear in each gap. Its sign also has a physical meaning: it indicates the direction of pumping of the end modes if the system is changed along the path. The Zak phase itself, however, is not without meaning. A value of $\tilde\sZ_1$ around $\pi/2$ ($3\pi/2$) indicates the edge states in the first gap will appear in the right (left) end of the ladder, due to its relation to the Wannier centers, as explained in Appendix \ref{ap:chern}.

The situation is completely analogous to the trimer chain case \cite{Martinez2019}. In Fig. \ref{fig:Chern} (d), the spectrum with periodic (light blue background) and open boundary conditions (lines) for the $\Delta$-CSSH ladder in subfigure (c) is represented along the path, for $L=16$. We can appreciate the non-protected end states, in green, moving around in their gaps, as dictated by their Chern number. The protected end states in the central gap, in red, stay constant at zero energy, as their Zak phase of $\pi$ dictates.

In the $\rhomb$-CSSH ladder, $\tilde C_1$ and $\tilde C_3$ are also nontrivial, while $\tilde C_2 = 0$. As discussed in Section \ref{sec:Rhomb}, one of the two outer gaps undergoes a topological phase transition as $m$ increases [see Fig. \ref{fig:Zak} (d)], across which the Zak phase jumps discretely but the number of end states remains constant. The end of the ladder in which the end state is located changes accordingly, from the right to the left end. This is accompanied by a change in the Chern number associated with that gap. If $\theta <\pi/4$, $\tilde C_1=-1$ if $-\pi<\phi<\pi \mod 4\pi$, where no phase transition is present in the first gap. For other values of $\phi$, $\tilde C_1$ changes from $-1$ to $1$ as $m$ increases and transverses its critical value. All signs are reversed if $\theta > \pi/2$. The behavior of $\tilde C_3$ is completely analogous, but the phase transitions are now only present if $-\pi<\phi<\pi \mod 4\pi$, and all signs are reversed.

\subsection{CSSH ladders with an odd number of rungs}

The discussion above assumes systems with an even number of rungs, that is, an integer number $N$ of unit cells. We now briefly discuss now the case when the number of rungs $L$ is odd. This means half a unit cell is left unpaired, either in the left or the right end of the ladder. If this choice of boundaries broke the protecting symmetries, the central gap edge states would not be protected anymore. However, all the chiral symmetries of the systems are preserved under this choice of boundaries, implying that, when present, the edge states in the middle gap will be chiral and protected. In the cases where the topology is protected by $\sX_C$ or $\sX_{\subrhomb}$, the states appear in the half unit cell when expected, and in the $\rhomb$- and $\Delta$-CSSH ladders they take an analogous form to their counterparts on the other end, something that is not true in the even-runged ladders.
This is a consequence of the presence of a global inversion symmetry, which will make the left and right states degenerate in the thermodynamic limit, even in the $m\ne 0$, $\phi \ne 0,\pi$ case.

In the cases where the topology is protected by $\sX_S$, the SSH-like chiral operator, topological states appear in the half unit cell when the Zak phase is trivial, and no states appear when it is topological. This is a well-known phenomenon in the SSH chain \cite{Ganeshan2013,Bello2016,Mei2018}, and can be easily understood in terms of the lattice geometry. In the D-class topological $\Delta$-CSSH, the states are protected by the PH symmetry $\sC_C$, which is not broken either in odd-runged ladders, and zero modes appear in the half unit cells. For a discussion of the $m=0$ states in even and odd-runged ladders, see Appendix \ref{ap:edge}.

Finally, the non-protected states in the first and third gaps of the $\rhomb$- and $\Delta$-CSSH ladders can also appear or disappear in the same way as SSH-type zero modes. Depending on the choice of boundaries, both left and right non-protected modes can be found in both ladders at the same time.

\section{Experimental implementations} \label{sec:exp}

Our work considers the single-particle case only, and so it is applicable for both bosonic and fermionic particles. As mentioned above, the Creutz ladder has been implemented in a cold atom system using orbital-momentum coupling \cite{Kang2018,HyounKang2020} and can be simulated in a specific regime \cite{Mukherjee2018} in a rhombus chain photonic lattice \cite{Mukherjee2018,Kremer2020,Jorg2020}. Recently, a Creutz ladder plaquette was also implemented in a superconducting system \cite{Hung2021}. Some other theoretical proposals have been made to implement the Creutz ladder in cold atom systems, either by using two hyperfine states as the legs of the ladder \cite{Piga2017} or two zigzag chains \cite{Sun2017}. A system of helical coupled-resonator waveguides \cite{Han2020} also produces the Creutz ladder as an effective model. By changing the tunneling amplitudes in the appropriate way, something particularly tunable in the mentioned systems, these setups can be employed to implement the CSSH ladders, under certain constraints. The easiest model to implement is probably the $\hgls$-CSSH ladder, given that the staggering of its amplitudes follows a simple SSH-like pattern. 

\begin{figure}[!htbp] 
    \includegraphics[width=\linewidth,trim=10 10 10 10,clip]{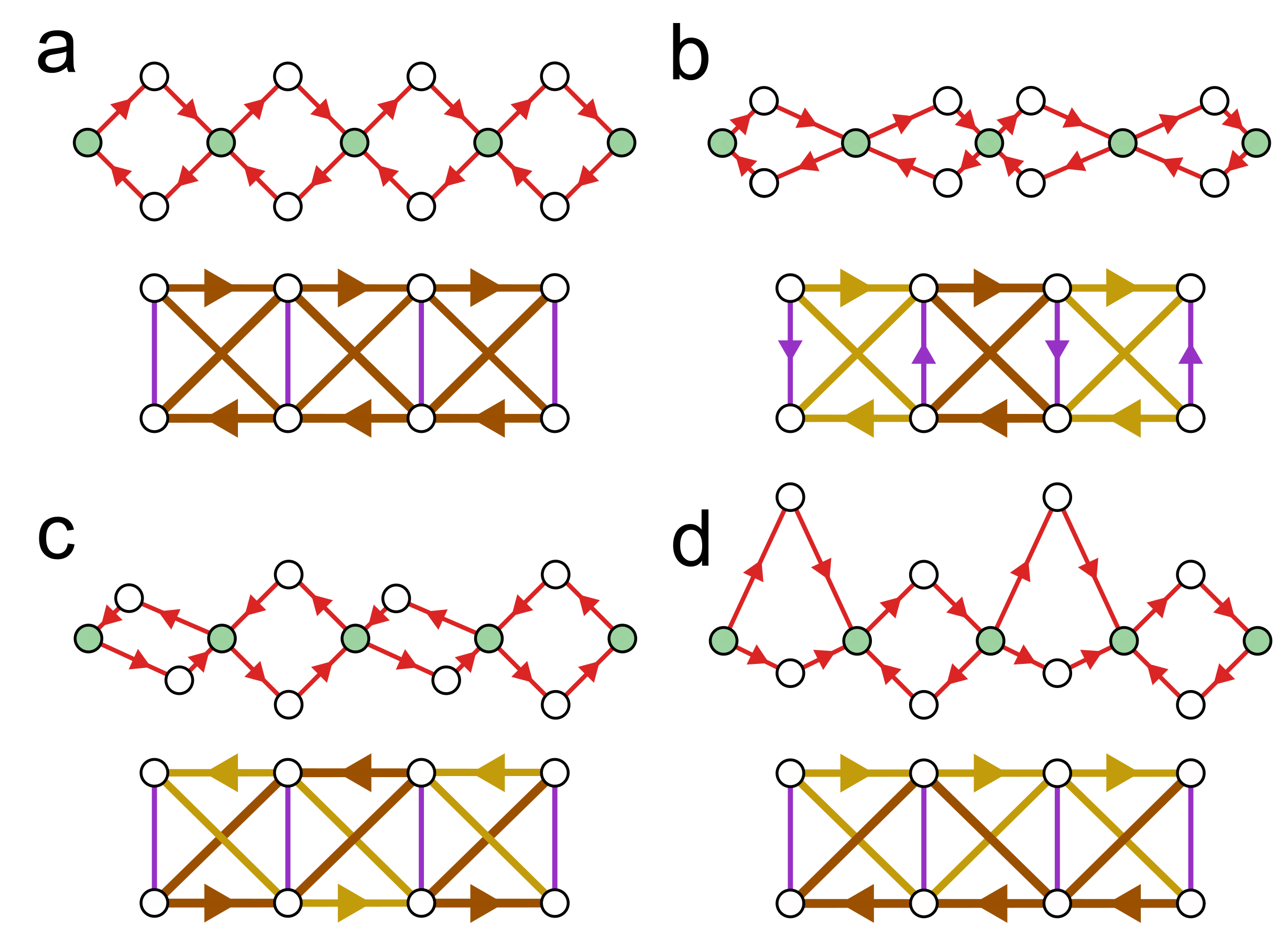}
    \caption{Photonic waveguide lattice diagrams for the CSSH ladders. (a) Photonic rhombus chain with a large energy offset in the middle sites (green). Its low-energy effective Hamiltonian is equivalent to the Creutz ladder \cite{Mukherjee2018}. (b) By deforming the rhombi as depicted, the $\hgls$-CSSH ladder can be implemented. For $\phi=\pi$, the rungless ($m=0$) regime is realized, while for other values of $\phi$, the vertical links are nonzero and acquire a phase. (c) If the odd rhombi are substituted by rhomboids, the $\rhomb$-CSSH ladder arises. As in (b), when $\phi=\pi$ the rungless model is achieved. When $\phi$ takes another value, $m$ is turned on. (d) If the odd rhombi are modified as pictured, a variant of the $\Delta$-CSSH ladder is obtained. It has some additional on-site energies that can be compensated to obtain the desired model. Like in the other ladders, the topological $\pi$-flux rungless case can be obtained, and other values of $\phi$ cause $m\ne 0$.}
    \label{fig:rhombi}
\end{figure}

Modified rhombus chains with a large chemical potential in its central sites can be used to implement the three CSSH models, following the Creutz ladder implementation presented in \cite{Mukherjee2018}. Their geometries are represented in Fig. \ref{fig:rhombi}. The $\pi$-flux $m=0$ case can be obtained directly, but the $\phi$ and $m$ degrees of freedom are not independent, and that restricts the regimes that can be obtained. For a detailed explanation and the relevant effective Hamiltonians, refer to Appendix \ref{ap:rhombi}. This general scheme can be used to implement the CSSH ladders in a photonic lattice or other systems, like ultracold atoms or superconducting circuits.

In particular, the tunneling amplitudes in cold atom implementations \cite{Sun2017,Piga2017,Kang2018,HyounKang2020} are related to the intensity of the lasers that induce the Raman-assisted transitions. This provides an easy way to independently tune each amplitude and obtain any of the CSSH ladders. In some proposals, like for instance \cite{Sun2017}, the $\phi$ and $m$ parameters can be tuned independently, and this would allow for the implementation of the D class TI or the rungless topological phase with any $\phi$.

Four-site versions ($L=2$) of all $m=0$ CSSH models can be implemented in the superconducting system in \cite{Hung2021}, by changing the appropriate tunneling amplitudes. It is worth noting that, in this limit, the $\hgls$-CSSH ladder reduces to a Creutz-style plaquette (which, in turn, can be rearranged as a square plaquette with a magnetic flux of $2\phi$), and the $\rhomb$ and $\Delta$-CSSH ladders become identical to each other. These systems, although small, still have states at exactly zero energy, which are protected by the same symmetries as those in the longer models.

The complex hopping amplitudes, which constitute one of the main difficulties in implementing these models, a periodic on-site energy driving in a waveguide lattice \cite{Mukherjee2018}, and it appears naturally as a result of the two-photon transition in the cold atom implementation \cite{HyounKang2020}, and of the pump that couples the modes in the superconducting scheme \cite{Hung2021}.

Additionally, given that the fermionic Creutz model is equivalent to a spin-$1/2$ particle hopping in a chain of sites in a specific gauge field \cite{Li2015B,Piga2017,Barbarino2019}, different fermionic implementations could also be considered for both the Creutz and the CSSH ladders, including fermionic cold atoms, semiconducting quantum dots or trapped ions, provided that the appropriate gauge field and different tunneling amplitudes and spin-orbit couplings can be established. 

\section{Conclusions}\label{sec:conclusion}

In this work, we propose and analyze a family of quasi-1D topological insulators, the CSSH ladders, with rich topological phase diagrams and flat-band regimes due to Aharonov-Bohm caging. 
Each of the models has some peculiar properties. The hourglass CSSH ladder can exhibit Creutz-like and SSH-like topological zero modes. The latter arise in the absence of a magnetic field, in contrast with the Creutz ladder case, where the $\phi=0$ model is trivial. The rhomboid CSSH ladder has a tunable chiral symmetry, which can be used to predict which will be the best quasi-1D model to use in each experimental setup to increase the protection against disorder. And, finally, the $\pi$-flux delta CSSH ladder is a quasi-1D topological material in the D class with no superconducting term, something rather unusual. By changing its parameters, this topological phase can be adiabatically connected with a BDI ($\phi=\pi$,$m=0$) and an AIII phase ($\phi\ne 0,\pi$,$m=0$), without losing the protection of the edge states. This can be of interest to explore the differences between 1D topological symmetry classes in a simple model.

The wide variety of topological states presented here are also tunable by changing the $m$, $\phi$ or $\theta$ parameters, which provides a high degree of control. They also have the robustness implicit to topological protection, which is still present in small, experimentally accessible systems in the rungless case.
For these reasons, they are highly suitable to be used in the fields of quantum information and computation. For instance, they could be used in a quantum memory, as part of a symmetry-protected qubit architecture, or as a means of transferring particles or states between the two ends of the ladder in a robust way. Additionally, the topological end modes in the rungless CSSH ladders are pinned at exactly zero energy, even for ladders only a few sites long, thanks to AB localization. This feature is inherited from the Creutz ladder, and makes a wide variety of topologically protected zero modes experimentally available, even in small systems. An interesting follow-up to this research would be to explore the potential of these topological zero modes in different quantum information tasks, and compare them to other topological systems.
Other intriguing topic is the study of the non-protected states in the rhomboid and delta CSSH ladder, which do not follow the bulk-boundary correspondence and are instead related to a Chern number. Previously reported on the trimer chain, they are reminiscent of topological states in quasicrystals. Their generalization to higher dimensions may be a promising topic for future research.
Finally, the applications of the wide variety of topological pumps that are possible in these models, some of them present in the Creutz ladder \cite{Sun2017} and others in the trimer chain \cite{Ke2016}, also constitute an interesting subject to explore further.

Finally, we propose a setup to implement the CSSH ladders in a photonic lattice system as an effective low-energy model in three modified rhombus chains. The large number of flat-band systems recently implemented in this kind of environment show that these lattices can be realized with the current technology. Given that our study applies to both bosons and fermions, the CSSH ladders can be realized in many other platforms, such as cold atoms, semiconducting quantum dots or superconducting circuits.

\begin{acknowledgements}
We would like to thank {\'A}lvaro G{\'o}mez-Le{\'o}n and Luca Leonforte for fruitful discussions. C. E. Creffield was supported by Spain’s MINECO through Grant No. FIS2017-84368-P. G. Platero and J. Zurita were supported by Spain’s MINECO through Grants No. MAT2017- 86717-P and PID2020-117787GB-I00 and by CSIC Research Platform PTI-001. J. Zurita recognizes the FPU program FPU19/03575.
\end{acknowledgements}

\appendix

\section{Topological invariants}\label{ap:zak}

To probe the topology of the CSSH ladders, we have used the Zak phase, calculated numerically using bulk Hamiltonian \eqref{eq:Hk}. In this Appendix, we specify some additional details about this calculation, which are not often discussed in the literature.

In the strong topological insulator phases of the models (that is, when $\phi=\pi \mod 2\pi$ or $m=0$), they belong to the symmetry classes BDI, AIII and D, which can harbour topological phases in 1D. The topological invariant in classes BDI and AIII is the winding number $\nu$, of type $\dZ$. In class D, the invariant, $\sQ$, is related to the Pfaffian of the bulk Hamiltonian, of type $\dZ_2$.

However, these topological invariants satisfy the following relations to the Zak phase $\sZ$ \cite{Sun2017,Velasco2019,Budich2013,Li2016}:
\begin{align}
    &\nu=\sZ/\pi \mod 2 \label{eq:nuZak}\\
    &\sQ=\sZ/\pi,       \label{eq:QZak}
\end{align}
provided spatial inversion symmetry is present and the center of the unit cell is chosen as the origin of coordinates.

Even in the cases where inversion symmetry is not present, the Zak phase is still a good topological invariant for protected zero modes, as we explain in the main text, and can also give us information about the non-conventional end states, related to a Chern number instead, that appear in the $\rhomb$- and $\Delta$-CSSH ladders (see subsection \ref{sec:DStates} and Appendix \ref{ap:chern}).

In the CSSH ladders, two distinct nontrivial phases appear, with $\nu=\pm 1$, corresponding to $0<\phi<2\pi$ and $-2\pi<\phi<0$ respectively, taken $\!\! \mod 4\pi$. These two phases cannot be distinguished with the Zak phase alone. They can be characterized by the chirality of their edge states \cite{Li2015,Sun2017}, just like their counterparts in the Creutz ladder, to which they can be connected adiabatically. Apart from this detail, our work focuses between the difference between trivial and nontrivial phases, to which the Zak phase is sensitive.

To calculate the Zak phase numerically, we have used the following formula, which can be obtained from its definition \eqref{eq:Zak} by discretizing momentum space and using the Taylor formula for the logarithm \cite{Resta1998,Vanderbilt}:
\begin{equation}\label{eq:ZakNum}
    \sZ = -\textrm{Im} \enskip \textrm{log} \prod_{k\in BZ} \braket{u_k^i|u^i_{k+\delta k}}
\end{equation}
where $\{\ket{u_k^i}\}$ is the $i$-th eigenstate of the bulk Hamiltonian $\sH(k)$ of the system. This expression will be valid as long as the step $\delta k$ in momentum space is small enough. It is independent of the gauge chosen for the eigenvectors $\{\ket{u_k^i}\}$, something necessary for numerical calculations.

In the convention we use for the bulk Hamiltonian, which is sometimes called Basis I in the literature \cite{Bena2009,Cayssol2021}, the Zak phase calculated using \eqref{eq:ZakNum} does not depend on the positions of the orbitals inside the unit cell, given that the unit cell number $n$ is treated as the spatial coordinate, and $\alpha$ as an internal degree of freedom. Another convention exists, Basis II, in which these positions would be taken into account in the bulk Hamiltonian, because the spatial coordinate is made to depend on both $n$ and $\alpha$. The version of the Zak phase obtained in this basis carries more information than in Basis I, because it depends on the position of the orbitals. For example, the Wannier centers calculated with it will be sensitive to the distribution of the wavefunctions inside each unit cell, while the Zak phase in Basis I will not. These distinctions have been noted in recent works, and some authors maintain that the name ``Zak phase'' should be reserved for the quantity that contains information about the positions of the orbitals \cite{Cayssol2021, Fuchs2021}. 
However, we keep the terminology for its Basis I counterpart, in agreement with many works in the literature \cite{Delplace2011,Longhi2018,Li2015,Bello2019,Aihara2020,Perez-Gonzalez2019a,Perez-Gonzalez2019b,Jiao2021,Cooper2019} and the reference text by Asb\'{o}th et al. \cite{Asboth2015}. When calculating the Zak phase in Basis II or in the continuum, an additional non-homogeneous phase factor has to be included in the last ket of the closed loop in expression \eqref{eq:ZakNum} \cite{Vanderbilt}.

Even though the Zak phases in Bases I and II are two distinct quantities, both are equally valid to study the topology of 1D materials. Basis I seems to be used more often in the literature, probably because its calculations are more straight-forward \cite{Asboth2015,Delplace2011,Longhi2018,Li2015,Bello2019,Aihara2020,Perez-Gonzalez2019a,Perez-Gonzalez2019b,Jiao2021,Cooper2019}. Works on the experimental detection of the Zak phase have used both Basis I \cite{Jiao2021} and Basis II \cite{Atala2013}.

Finally, we want to note an important point. The Zak phase, in both bases, depends on the choice of the origin of coordinates. This happens because that choice is related to a gauge transformation on the Bloch states $\{\ket{u_k^i}\}$ \cite{Vanderbilt,Marzari1997}. In particular, the Bloch states at $k=\pm\pi/2$ can have, in general, different phases. The expression we use, equation \eqref{eq:ZakNum}, corresponds to the Zak phase calculated using a periodic gauge for the cell-periodic Bloch states, where both states are equal. In Basis I, the choice of coordinates $x_n = na_0$ (where $a_0$ is the lattice constant, taken to be 1 in the main text), which we use, provides this type of gauge, and also allows us to use expressions \eqref{eq:nuZak} and \eqref{eq:QZak}.

\section{Full list of symmetries}\label{ap:syms}
The full list of symmetries defined in the text is detailed below. The conventional, untilded symmetries act on $\sH(k)$, and are written in the standard internal basis. As usual, they act on the spatial coordinate by leaving it unchanged, if unitary, or with a complex conjugation, if antiunitary. The hidden symmetries are represented by tilded operators, which act on the full Hilbert space and are expressed in the standard basis.

The symmetries of the SSH chain Hamiltonian \eqref{eq:SSH}, expressed in the $\{\k{I},\k{I\!I}\}$ basis, are $\sC^{(2)}_S = \sigma_z \sK$,
    $\sT^{(2)}_S = \sK$ and $\sX^{(2)}_S = \sigma_z$,
where the superindex is a reminder of the dimension of the matrices and $\sK$ is complex conjugation.

The $m=0$ hourglass and rhomboid CSSH ladders satisfy a chiral symmetry of this type:
\begin{align}
    \sX_S = \begin{pmatrix}
    \uno_2 & 0\\
    0 & -\uno_2    
    \end{pmatrix},
\end{align}
in the basis $\{\k{A},\k{B},\k{C},\k{D}\}$, where $\uno_2$ is the $2\times 2$ identity matrix. The $m=0$ Creutz ladder satisfies the related hidden symmetry:
\begin{equation}
    \tilde\sX_{S} = \diag\nolimits_{2L}(1,1,-1,-1,1,1,-1,-1,\ldots),
\end{equation}
in the basis $\{\k{1,a},\k{2,b},\ldots,\k{L,b}\}$.

The Creutz ladder Hamiltonian with $\phi=\pi$ satisfies the symmetries $\sC^{(2)}_C = \sigma_z \sK$, $\sT^{(2)}_C = \sigma_x\sK$ and $\sX^{(2)}_C = i\sigma_y$,
written in the basis $\{\k{a},\k{b}\}$.

The symmetry $\sT^{(2)}_C$ flips the upper and lower legs of the ladder, something unexpected in the bosonic case \cite{Hugel2014}. It is, however, a more natural TR transformation if we consider the two legs of the ladder as the up and down components of a spin-$1/2$ particle, something we can do if we take $c^{(\dagger)}_{n,\alpha}$ to be fermionic operators. This approach has been explored in \cite{Sticlet2014,Hugel2014,Li2015B,Barbarino2019,Creffield2020}. However, a fermionic TR symmetry usually squares to $-\uno$, not $\uno$ as is the case, so it is still non-conventional in that regard. In any case, $\sT^{(2)}_C$ behaves like a regular TR symmetry in all cases as far as topology is concerned, so we keep the usual nomenclature for its operator and the associated symmetry classes. The same can be said about many of the transformations below, which have a non-conventional form but fulfill their role as chiral, PH or TR symmetries.

We say then that the $\pi$-flux Creutz ladder belongs to the BDI class \cite{Tovmasyan2013,Sticlet2014,Piga2017,Gholizadeh2018,Alaeian2019}. Other authors do not consider the mentioned TR and PH symmetries, and thus classify the system as a member of the AIII class, with only chiral symmetry \cite{Li2015,Jafari2019}. Given that the chiral symmetry is enough to protect the topological states, both classifications are valid.

Even though $\sC^{(2)}_S$ and $\sC^{(2)}_C$ have the same matrix expression, they have different generalizations in the CSSH ladders due to the different degrees of freedom in which they act. 
The analogues of the mentioned symmetries in the CSSH ladders are (see Section \ref{sec:topo}):

\begin{align}
    &\sC_C = \begin{pmatrix}
    \sigma_z & 0 \\
    0 & \sigma_z
    \end{pmatrix} \sK\\
    &\sT_C = \begin{pmatrix}
    \sigma_x & 0 \\
    0 & \sigma_x
    \end{pmatrix}\sK\\
    &\sX_C = i\begin{pmatrix}
    \sigma_y & 0 \\
    0 & \sigma_y
    \end{pmatrix}.
\end{align}

There are also some composite symmetries, formed by the composition of an SSH-type and a Creutz-type symmetry (hence the $SC$ subindex):
\begin{align}
     &\sC_{SC} =\sX_S \sT_C^{-1} =\begin{pmatrix}
    \sigma_x & 0\\
    0 & -\sigma_x
    \end{pmatrix} \sK\\
    &\sT_{SC} =\sC_C^{-1} \sX_S =\begin{pmatrix}
    \sigma_z & 0\\
    0 & -\sigma_z
    \end{pmatrix} \sK,
\end{align}
expressed in the internal CSSH ladder basis.

The $m=0$ Creutz ladder satisfies the following hidden composite PH symmetry:
\begin{equation}
    \tilde \sC_{SC} =\diag\nolimits_{L}(\sigma_x,-\sigma_x,\sigma_x,-\sigma_x,\ldots) \sK,
\end{equation}
in the basis $\{\k{1,a},\k{2,b},\ldots,\k{L,b}\}$.

Some symmetries of the  $\pi$-flux rhomboid CSSH ladder have a unique form:
\begin{align}
    &\sT_{\subrhomb} = \begin{pmatrix}
    -is_\varth & c_\varth & 0 & 0\\
    c_\varth & -i s_\varth & 0 & 0\\
    0 & 0 & 0 & 1\\
    0 & 0 & 1 & 0
    \end{pmatrix} \sK \label{eq:TRhomb}\\
    &\sX_{\subrhomb} = \begin{pmatrix}
    is_\varth & c_\varth & 0 & 0\\
    -c_\varth & -i s_\varth & 0 & 0\\
    0 & 0 & 0 & 1\\
    0 & 0 & -1 & 0
    \end{pmatrix}
\end{align}
where
\begin{align}
    s_\varth = \sin \varth = \frac{J'^2-J^2}{J'^2+J^2}\\
    c_\varth = \cos \varth = \frac{2J'J}{J'^2+J^2},
\end{align}
and with $\varth=2\arctan(\cos 2\theta)$. These symmetries reduce to $\sT_C$ and $\sX_C$ when $J=J' (\varth=0)$, because the model becomes a Creutz ladder. On the other hand, when $\varth=\pm \pi/2$, which corresponds to $J=0$ and $J'=0$ respectively, we obtain the orthogonal dimer chain, with the following symmetries:
\begin{align}
    &\left.\sT_{\subrhomb}\right|_{\varth=\pm \pi/2} = \begin{pmatrix}
    \mp i \uno_2 & 0 \\
    0 & \sigma_x
    \end{pmatrix} \sK\\
    &\left.\sX_{\subrhomb}\right|_{\varth=\pm \pi/2} = \begin{pmatrix}
    \mp i \sigma_z & 0 \\
    0 & i\sigma_y
    \end{pmatrix},
\end{align}
in addition to the PH symmetry $\sC_C$, present for all $\rhomb$-CSSH ladders with $\phi=\pi$.

The chiral and PH symmetries protect the zero modes in these models, as long as they are not broken. These symmetries are usually broken by on-site energy disorder. In the Creutz limit, however, the chiral symmetry can withstand any on-site energy configuration, as long as the energies in the upper leg of the ladder have the opposite value as those in the lower leg. This will also make it more robust against general on-site energy disorder, given that some of the disordered configurations will randomly approach this case. The chiral symmetry in the orthogonal dimer chain, however, only resists symmetric on-site energies in the $C,D$ sites.

We now show this analytically. We can quantify the degree of symmetry breaking in an open system with on-site energies in the odd rungs, $\epsilon_A,B$, with the following quantities:
\begin{align}
    &x = \|{\tilde\sX_{\subrhomb} \sH_{\subrhomb} \tilde\sX_{\subrhomb}^\dagger + \sH}\|\\
     &t = \|{\tilde\sC_{C} \sH_{\subrhomb} \tilde\sC_{C}^{-1} + \sH}\|,
\end{align}
where $\|{\cdot}\|$ is the Frobenius norm, and the Hamiltonian $\sH_{\subrhomb}$, as well as the tilded operators, acts on the full real space. The results, in the Creutz and orthogonal dimer chain limits, are, respectively,
\begin{align}
    &x_C=2\sqrt{2}|\epsilon_A + \epsilon_B| \\
    &x_{OD} = 4\sqrt{\epsilon_A^2 + \epsilon_B^2}
\end{align}
while $t=x_{OD}$ for all values of $\theta$. Thus, $x_{OD}>0$ for any value of the energies, while $x_C$ is zero as long as $\epsilon_A = -\epsilon_B$.

The TR symmetry for the $\rhomb$-CSSH ladder with a general value of $\phi$ has a more complicated form:
\begin{align}
    &\sT'_{\subrhomb} = \begin{pmatrix}
    S(\theta,\phi) & C(\theta,\phi) & 0 & 0\\
    C(\theta,\phi) & S(\theta,\phi) & 0 & 0\\
    0 & 0 & 0 & 1\\
    0 & 0 & 1 & 0
    \end{pmatrix} \sK, \label{eq:TTun}
\end{align}
with
\begin{align}
    &S(\theta,\phi) = e^{i\phi/2} \frac{J^2 - J'^2}{J^2 -e^{i\phi} J'^2}\\
    &C(\theta,\phi) = (1-e^{i\phi})\frac{JJ'}{J^2-e^{i\phi}J'^2},
\end{align}
which does not depend on $\xi$, only on $\theta$ (as mentioned in the text, $J=\xi\sin^2\theta$ and $J'=\xi\cos^2\theta$). Equation \eqref{eq:TRhomb} is actually a special case of this operator, with $\phi=\pi$.
When $m=0$, the system also has the chiral symmetry $\sX_S$ and a PH symmetry represented by:
\begin{align}
    &\sC'_{\subrhomb} \!=\! \sX_S \sT_{\subrhomb}^{\prime -1}\! =\! \begin{pmatrix}
    S(\theta,\phi) & C(\theta,\phi) & 0 & 0\\
    C(\theta,\phi) & S(\theta,\phi) & 0 & 0\\
    0 & 0 & 0 & -1\\
    0 & 0 & -1 & 0
    \end{pmatrix} \sK.
\end{align}

The SSH chain with complex amplitudes that appears in one of the subspaces of the rungless $\pi$-flux $\rhomb$-CSSH ladder has the symmetries $\sC^{(2)}_W =\sigma_x\sK$, $\sT^{(2)}_W =\sigma_z \sK$ and $\sX^{(2)}_W = \sigma_z$.

The unitary commuting symmetries that allow us to reduce the Hilbert space into two subspaces are:
\begin{align}
    &U = \begin{pmatrix}
        \sigma_x & 0\\
        0 & \sigma_x
    \end{pmatrix} \\
    &V = \begin{pmatrix}
        i\sigma_y & 0\\
        0 & -i\sigma_y
    \end{pmatrix} \\
    &W = \begin{pmatrix}
        i s_\varth & c_\varth & 0 & 0\\
        -c_\varth & -i s_\varth & 0 & 0\\
        0 & 0 & 0 & -1\\
        0 & 0 & 1 & 0
    \end{pmatrix},
\end{align}
present in the $\hgls$-CSSH ladder with $\phi=0$, in the $\hgls$-CSSH ladder with $\phi=\pi$ and $m=0$, and in the $\rhomb$-CSSH ladder with $\phi=\pi$ and $m=0$, respectively.

In the $\rhomb$ and $\Delta$-CSSH $\phi=0$ cases, the Hamitonian acquires a TR symmetry, $\sT_S=\sK$, and when $m=0$ it also has the PH and chiral symmetries $\sC_S=\sigma_z\otimes \uno \sK$ and $\sX_S=\sigma_z\otimes \uno$ (defined above). These are the four-site counterparts of the SSH symmetries. However, these systems are topologically trivial. A doubly degenerate flat band appears at zero energy when $m=0$, and two dispersive bands are located over and under it, touching it if $\theta=\pi/4$ in the $\rhomb$-CSSH ladder, and for all values of $\theta$ in the $\Delta$ ladder. When $m$ is turned on, one flat band remains at $E=m$, and the other three bands are dispersive. Non-protected states appear in the lower gap in the rhomboid ladder.

Finally, the inversion symmetry of the SSH, Creutz and $\hgls$-CSSH ladder with periodic boundary conditions can be expressed as
\begin{equation}
    \tilde \sI=\diag\nolimits_{M}^{-1}(1\ldots 1),
\end{equation}
where $M$ is the total number of sites in the system and $\sI=\diag_{\ell}^{-1}(a,\ldots,z)$ is the $\ell\times\ell$ antidiagonal matrix with elements $\sI_{1,\ell}=a,\ldots,\sI_{\ell,1}=z$.

In the $\rhomb$ and $\Delta$-CSSH ladders with periodic boundary conditions, the inversion symmetry can be represented by the operators:
\begin{align}
    &\tilde\sI_{\subrhomb} = \left(\begin{array}{c|c}
    \sI_{6} &0 \\ 
    \hline
        0  & \sI_{4N-6}
    \end{array}\right)\\
    &\tilde\sI_{\Delta} = \left(\begin{array}{c|c}
    \sJ_{6} &0 \\ 
    \hline
        0  & \sJ_{4N-6}
    \end{array}\right),
\end{align}
respectively, where $\sI_{n}=\diag\nolimits^{-1}_{n}(1,\ldots,1)$ and $\sJ_{n}=\diag\nolimits^{-1}_{n/2}(\sigma_x,\ldots,\sigma_x)$.

\section{Form of the caged topological edge states}\label{ap:edge}
In all CSSH ladders, the topological edge states are completely localized in the two sites of the end rungs of the model when $m=0$ and $\phi\ne 0$. This localization has the same nature as the AB caging of the $\pi$-flux case, but for other values of $\phi$ it does not affect the bulk states. As discussed in the main text, the protected modes will be pinned at exactly zero energy for any $L>1$, making topological protection meaningful in systems with as few as 4 sites. The analytical form of the zero modes for the $m=0$ ladders with an even number of rungs are:

\begin{align}
    &\k{\sL_{\subhgls}} \!=\! \frac{\k{1,A}-e^{i\phi/2}\k{1,B}}{\sqrt{2}}\\
    &\k{\sL_{\subrhomb}} \!=\! \k{\sL_{\Delta}} \!=\! \frac{J\k{1,A}-J' e^{i\phi/2}\k{1,B}}{\sqrt{J^2 + J'^2}}\\   
    &\k{\sR_{\subhgls}} \!=\! \k{\sR_{\subrhomb}} \!=\! \k{\sR_{\Delta}} \!=\! \frac{\k{N,C}-e^{-i\phi/2}\k{N,D}}{\sqrt{2}}   
\end{align}

When the number of rungs is odd, we find half a unit cell in one end of the system. The edge states in that end, which we mark with a prime, are still protected, but their form changes to one analogous to the state in the other side, but with the opposite chirality. Their form is:
\begin{align}
    &\k{\sL'_{\subhgls}} \!=\! \k{\sL'_{\subrhomb}} \!=\! \k{\sL'_{\Delta}} \!=\! \frac{\k{1,A}-e^{i\phi/2}\k{1,B}}{\sqrt{2}} \\
    &\k{\sR'_{\subhgls}}  \!=\! \frac{\k{N,C}-e^{-i\phi/2}\k{N,D}}{\sqrt{2}}\\
    &\k{\sR'_{\subrhomb}} \!=\! \frac{J'\k{N,C}-J e^{-i\phi/2}\k{N,D}}{\sqrt{J^2 + J'^2}} \\   
    &\k{\sR'_{\Delta}} \!=\! \frac{J\k{N,C}-J' e^{-i\phi/2}\k{N,D}}{\sqrt{J^2 + J'^2}}
\end{align}
We can see the states in the hourglass CSSH ladder do not change with respect to the even-runged case.

Finally, the only non-protected edge states that can experience AB caging are those in the $\Delta$-CSSH ladder, and only for $m=0$ and $\phi=\pi$. This is because the other models do not have any non-protected states for that parameter choice (see Fig. \ref{fig:2DspectraU}). One of them is represented in Fig. \ref{fig:states} (j). They have the form:
\begin{align}
    \ket{\ell\pm} = \frac{J'\ket{1,A} + iJ\ket{1,B}}{\sqrt{2(J^2 + J'^2)}} 
    \mp \frac{i\ket{1,C} + \ket{1,D}}{2}
\end{align}

\section{Non-protected end states and Wannier functions} \label{ap:chern}

In 1D topological insulators, the Zak phase acts like a $\dZ_2$ topological invariant, which can also distinguish between the even and odd phases of a $\dZ$-type phase (inside each group, different phases can be identified by their winding number). At the same time, it is related with the 1D electric polarization. As shown by Zak in its original paper \cite{Zak1989}, the Zak phase of a band $\sZ_i$ is related to its Wannier center $q_i$, by the relation $\sZ_i = 2\pi q_i$. This point is calculated as the center of mass in the spatial coordinate of the Wannier functions of the band.

A Zak phase of zero is associated with a Wannier function which is centered in its unit cell. This corresponds to the trivial insulating phase. If the Zak phase takes a value of $\pi$, the Wannier function has shifted half a unit cell, and it is now spread between at least two contiguous unit cells. This is directly related with the appearance of topological end modes: if open boundary conditions are enforced, the two Wannier functions at the ends will each be missing half of their support states, and so they will form states outside the bands \cite{Aihara2020}. This is the situation in the SSH chain, the Creutz ladder, the $\hgls$-CSSH ladder and for some parameters in the $\rhomb$-CSSH ladder. It is most easily seen in the flat band limit of the ladders, where an eigenbasis of Wannier functions can be chosen, which only have support over a small number of sites (i.e., they are \textit{compact}). Each Wannier function spans four sites in these models, in a similar way to the Creutz ladder case \cite{Tovmasyan2013,Murad2018}. The hourglass CSSH ladder bulk eigenstates are:
\begin{align}
    &\ket{u_{\subhgls n}^\pm} = \frac{1}{2}\left[\ket{n,A} + i\ket{n,B} \mp \left(i\ket{n,C} + \ket{n,D} \right) \right]\nonumber\\
    &\ket{v_{\subhgls n}^\pm}\! =\! \frac{1}{2}\left[\ket{n\!\!+\!\!1,A} \!-\! i\ket{n\!\!+\!\!1,B} \!\pm\! \left(i\ket{n,C}\! -\! \ket{n,D} \right) \right]\nonumber
\end{align}
with $n=1,\ldots,N$ the cell coordinate.
The intracell (intercell) states $\ket{u_{\subhgls n}^\pm}$ ($ \ket{v_{\subhgls n}^\pm}$) form flat bands at $E_{u\subhgls}^\pm = \pm2J'$ ($E_{v\subhgls}^\pm = \pm 2J$). The rhomboid CSSH ladder bulk states are:
\begin{align}
    &\ket{u_{\subrhomb n}^\pm} = \frac{1}{\kappa}(J'\ket{n,A} + iJ\ket{n,B}) +\nonumber\\
    &\mp\frac{1}{2}(i\ket{n,C} + \ket{n,D})\nonumber\\
    &\ket{v_{\subrhomb n}^\pm} = \frac{1}{\kappa}(J\ket{n+1,A} - iJ'\ket{n+1,B}) +\nonumber\\
    &\pm\frac{1}{2}(i\ket{n,C} - \ket{n,D})\nonumber
\end{align}
with $n=1,\ldots,N$.
The positive energy inter- and intracell states are degenerate, with energy $E_+=\sqrt{2(J^2 + J'^2)}$. The degenerate negative energy states have energy  $E_-=-\sqrt{2(J^2 + J'^2)}$.

\begin{figure}[!htbp] 
    \includegraphics[width=\linewidth,trim=10 10 10 10,clip]{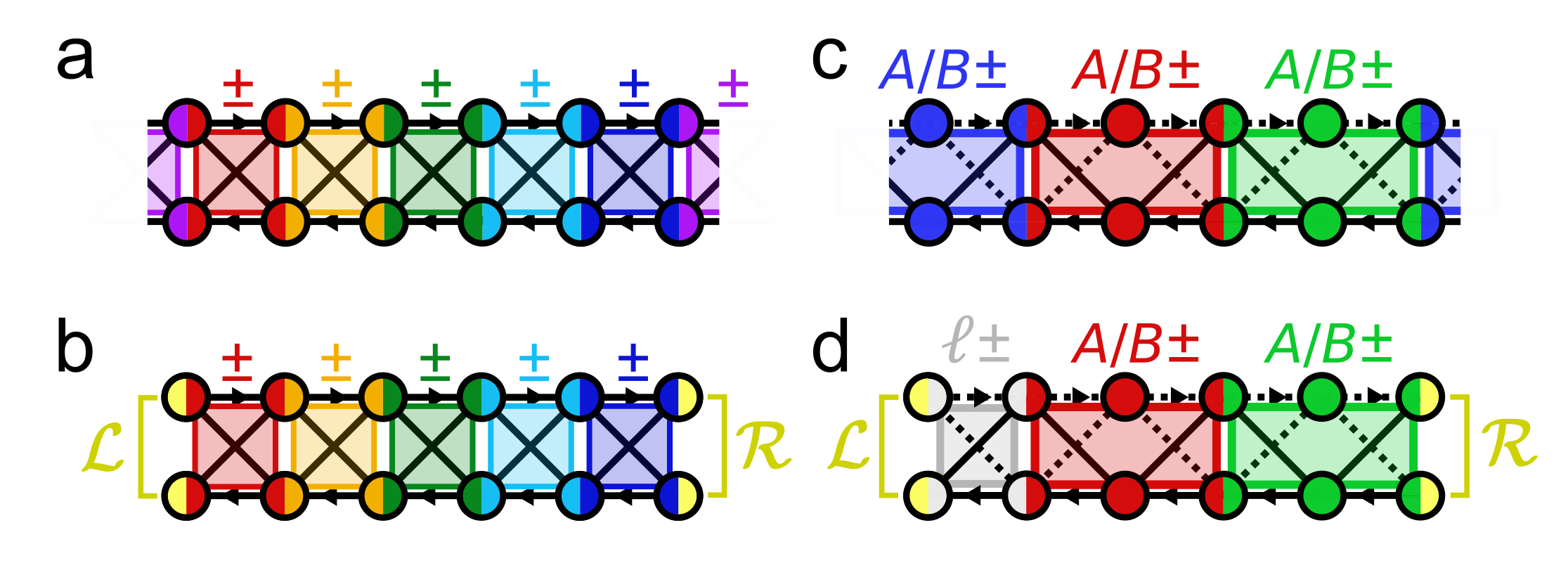}
    \caption{Wannier basis and topological edge states in the flat-band ladders. (a) Wannier states in the Creutz, hourglass and rhomboid CSSH ladders with periodic boundary conditions. Each pair of Wannier functions, marked with $\pm$, span four contiguous sites. (b) In the open system, two unpaired halves of Wannier states, one at each end, become the topological edge states. (c) In the periodic delta CSSH ladder, each group of four Wannier states, labelled by $A/B\pm$, span six sites in total. The whole ladder is covered by them. (d) In the open system, three modes in total are left unpaired on the left end of the ladder, and one on the right end. These become the two protected zero modes, $\ket{\sL},\ket{\sR}$, and the two non-protected end states $\ket{\ell\pm}$, which appear in the first and third gaps.}
    \label{fig:Wan}
\end{figure}

The intercell Wannier functions are centered on the boundary between unit cells, and so they are responsible for the topological Zak phase and the appearance of end modes. We demonstrate this in a schematic way in Fig. \ref{fig:Wan} (a). In the Creutz and $\hgls$-CSSH ladders, the bulk-boundary correspondence always holds, and the number of end states in a gap is predicted by the sum of the Zak phases of the bands below.

This is not the case for the $\rhomb$- and $\Delta$-CSSH ladders. An example can be seen in the flat band $\Delta$-CSSH ladder.
Its compact Wannier function eigenbasis is:
\begin{align}
    &\ket{w_{\Delta n}^{(A)\pm}} = \frac{1}{\sqrt{2}}\ket{n,A} \mp \frac{i}{\sqrt{8}}\left[\ket{n,C} - \ket{n-1,C}\right.\nonumber\\
    &\left.-i \left(\ket{n,D} + \ket{n-1,D} \right) \right]\nonumber \\
    &\ket{w_{\Delta n}^{(B)\pm}} = \frac{1}{\sqrt{2}}\ket{n,B} \pm \frac{i}{\sqrt{8}}\left[\ket{n,D} - \ket{n-1,D}\right.\nonumber\\
    &\left. +i \left(\ket{n,C} + \ket{n-1,C} \right) \right] \nonumber
\end{align}
with $n=1,\ldots,N$.
The flat bands spanned by the two former states have $E=\pm 2J' $, and those spanned by the two latter, $E = \pm 2J$.

These Wannier states span five sites each (located on three rungs), and are always centered on the odd rungs of the ladder, as we depict in Fig. \ref{fig:Wan} (c). Their centers are located at a position of $3/4$ in each unit cell (with a lattice constant of $a_0=1$ between equivalent sites). Given that $\sZ_1 = 3\pi/2$ there [see Fig. \ref{fig:Zak} (e)], the first-band Wannier center position coincides with $\sZ_1/2\pi$, as expected. In the bulk, all states are accounted for, each $A$ or $B$ site participates in two of the Wannier states, and each $C$ or $D$ site is part of four of them. When the system is opened (without breaking any unit cells), a single end mode $\ket{\sR}$ appears on the right side, in the same way than in the other ladders. However, in the left end, a total of 3 modes are left unpaired, which reorganize into a protected end state $\ket{\sL}$ in the middle gap and two unprotected topological states $\ket{\ell\pm}$, with support on the four first sites, in the first and third gaps [see Fig. \ref{fig:Wan} (d)]. Away from the flat band point, the Wannier functions are no longer energy eigenstates, but they can still cause the appearance of the end states can still appear. An analogous situation occurrs in the $\rhomb$-CSSH ladder, although not in its flat band limit, so the analysis of its Wannier functions would be more involved.




By referring to the $\Delta$-CSSH flat band case again, we can conclude that the topological states do not only appear because the compact Wannier functions are off-center, but also because they cannot fit inside one unit cell, and this phenomenon cannot be captured by the usual Zak phase. This situation is reminiscent to the results of a very recent work \cite{Nelson2021}, in which the authors propose a 3D topological material in which the topology is also not captured by the standard tenfold classification. The reason for this is that the maximally localized Wannier functions of the system cannot fit into a single primitive unit cell, and this causes the apparition of topological surface states. We believe the trimer chain and the delta CSSH ladder might be, in this sense, 1D analogues of this phenomenon, termed by the authors ``multicellular topology'', although further research is needed.


\section{Edge state protection}
\label{ap:prot}

The topological edge states in the second gap of the CSSH ladders, when present, are protected by the chiral and PH symmetries (or only by the PH symmetry, in the D class $\Delta$-CSSH ladder). These symmetries force the spectrum to be symmetric around zero, and will typically relate two linear combinations of the left and right states. These linear combinations will be energy eigenstates, and then they will typically have small, finite energies around zero, as long as hybridization of the left and right modes is allowed. However, if hybridization is inhibited, whether because the system is large enough or due to another mechanism (like AB caging), the left and right edge modes will also be energy eigenstates. The protecting symmetries will then force their energy to be exactly at zero. This is the sign of topological protection in a 1D system.

\begin{figure*}[!htbp] 
    \includegraphics[width=\linewidth,trim=2 2 2 2,clip]{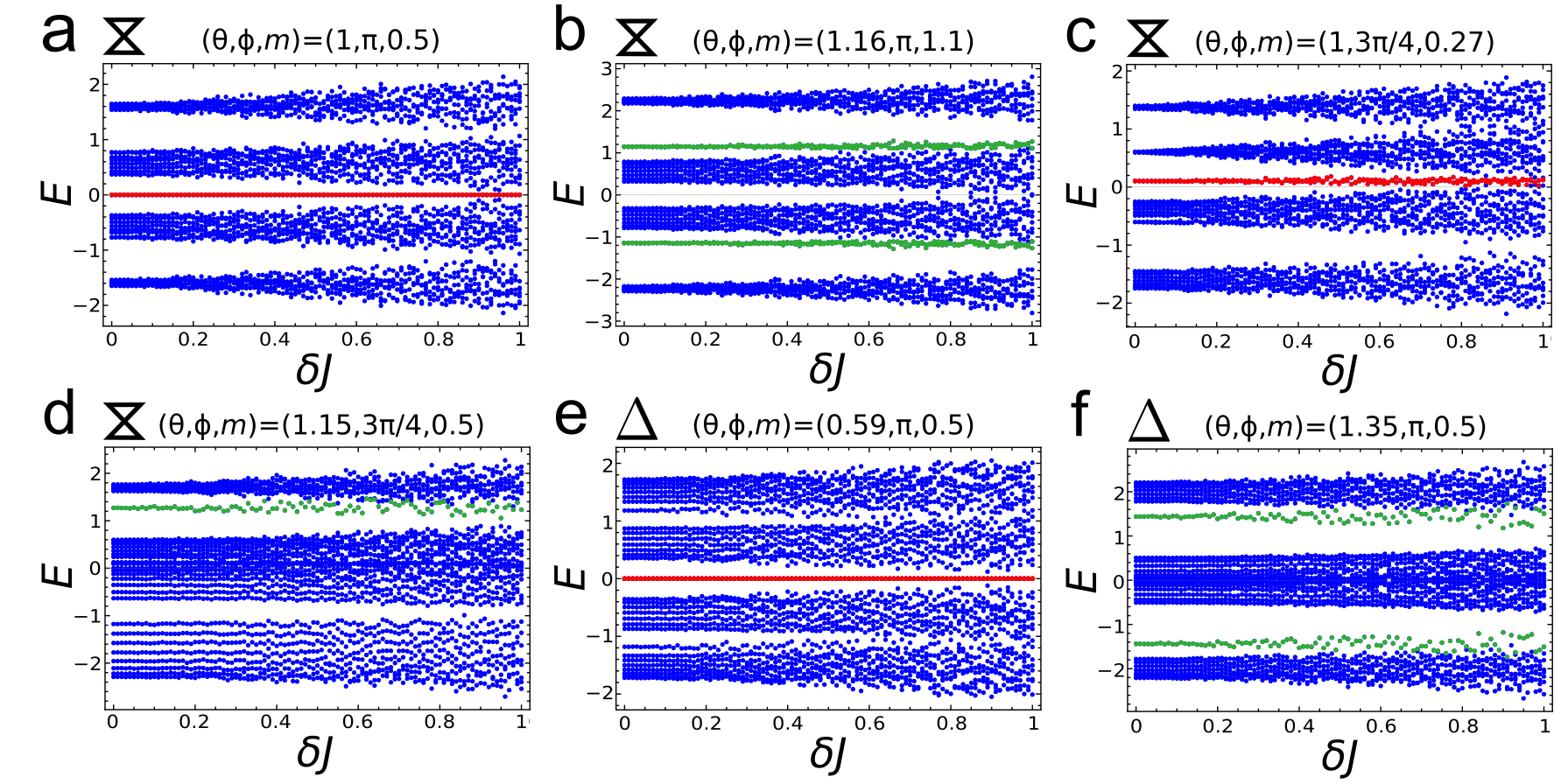}
    \caption{Spectra of hourglass (a-d) and delta (e-f) CSSH ladders with open boundary conditions as a function of the level of hopping amplitude disorder $\delta J$, for the parameters shown above each plot. All simulations use $L=16$, $\xi = 1$. The parameters were chosen so that the studied end mode in each case has a distance of $\Delta E = 0.35$ with the closest band, for the effects of disorder to be comparable. The end states in the second (first or third) gap are represented in red (green). (a) Protected zero modes. Their energy is fixed, even for large levels of disorder. (b) Non-protected edge states. For large levels of disorder, their energies drift up and down. (c,d) Edge states in a topological crystalline insulator, not protected against disorder. (e) Protected zero mode in the $\Delta$-CSSH ladder. (f) Non-protected left edge states $\ket{\ell \pm}$, which do not follow the usual bulk-boundary correspondence.}
    \label{fig:dis}
\end{figure*}

In Fig. \ref{fig:dis}, the effect of hopping disorder in the terms $t_i$ of the CSSH Hamiltonian \eqref{eq:CSSH} is studied. The disorder is modeled as:
\begin{equation}
    t_i = t_i^{c} + \delta J R,
\end{equation}
where $\delta J$ is the level of disorder, $t_i^c$ is the value of the amplitude in a clean system, and $R$ is a uniformly drawn random number between $-1/2$ and $1/2$, chosen individually for each term in the Hamiltonian and for each value of $\delta J$. The level of disorder ranges from zero to $\xi$ (with $\xi = 1$) to better appreciate its effects, but it must be noted that the level of disorder expected in a real experimental setup would be much smaller, on the order of $0.1\xi$ \cite{Lang2017,Mei2018,Boross2019}.

In Fig. \ref{fig:dis} (a) and (e), protected zero modes in the $\hgls$- and $\Delta$-CSSH ladders are represented. Their energy is fixed at exactly zero, even in the presence of intense disorder, because disorder in the hopping amplitudes of the system does not break the protecting symmetries.

If both the chiral and the PH symmetries are absent, or if we consider edge states that are not located in the gap around zero energy, then the situation changes. The absence of each of these conditions must be studied separately.

If both protecting symmetries are absent, then the system belongs to a symmetry class which cannot harbor topological phases in 1D, according to the tenfold classification. However, the presence of crystalline symmetries can expand and modify this classification, allowing for the presence of degenerate topological end modes. The system is then a topological crystalline insulator. As described in the text, this is the case in the Creutz and $\hgls$-CSSH ladders with $\phi \ne 0,\pi$ and $m\ne 0$: the central topological end modes of the spectrum are degenerate due to inversion symmetry, at an energy value different than zero. However, inversion symmetry will be broken by any physically relevant disorder present in the system, and so we refrain from categorizing these states as ``protected'' in this sense. States of this kind are depicted in Fig. \ref{fig:dis} (c). Their energy stays close to its value in a clean system, but it starts drifting up and down as the disorder strength increases.

The second kind of non-protected topological states are those in the first and third gaps of the system. The protecting symmetries can only cause their energies to have opposite values, not to be fixed at a given value. For this reason, they are not prevented from drifting into the bulk bands when a high level of disorder is present.

In Fig. \ref{fig:dis} (b), the response of non-protected states in the first and third gaps of the $\hgls$-CSSH ladder spectrum can be seen for different values of $\delta J$. As disorder increases, their energies start fluctuating around their value in the clean system. This also happens in the topological crystalline insulator case, Fig. \ref{fig:dis} (d), and for the left edge modes in the $\Delta$-CSSH ladder which do not follow the usual bulk-boundary correspondence, Fig. \ref{fig:dis} (f). In these two last cases, the fluctuations are more pronounced than in Fig. \ref{fig:dis} (b), probably due to the absence of PH and chiral symmetries in (d), and the different nature of the $\ket{\ell \pm}$ states in (f). In all the analyzed models, except in the topological crystalline insulator phase, the spectrum is still symmetric around zero energy, because the PH and chiral symmetries are not broken. This means that when the state in the first gap drift upwards, the one in the third gap will drift downwards, and vice versa.

\section{Rhombus chain implementation} \label{ap:rhombi}
We consider a generalized rhombus chain Hamiltonian with a 6-site unit cell (see Fig. \ref{fig:rhombi}). We label the upper states $A$ and $C$, the lower states are $B$ and $D$, and the middle states are $X$ and $Z$:
\begin{align}
    &\sH_R = -\sum_n \left\{\vphantom{\frac{\eps}{2}} v_1^+ \kb{n,A}{n,X} + v_2^+ \kb{n,X}{n,B}\right.\nonumber\\
    &+v_3^+\kb{n,Z}{n,A}+v_4^+\kb{n,B}{n,Z} \nonumber\\
    &+w_1^+ \kb{n,C}{n,Z} + w_2^+ \kb{n,Z}{n,D}\nonumber\\
    &+w_3^+\kb{n+1,X}{n,C}+w_4^+\kb{n,D}{n+1,X}\nonumber\\
    &\left. + \frac{\epsilon}{2}\kb{n,X}{n,X} + \frac{\epsilon}{2}\kb{n,Z}{n,Z} + h.c.\right\} ,
\end{align}
where $v_i^+=v_i e^{i\phi/4}$ and $w_i^+=w_i e^{i\phi/4}$. The phases implement a magnetic flux of $\phi$ through each rhombus of the chain, and can be generated using Floquet driving \cite{Mukherjee2018}.

If we consider the regime where $\epsilon\gg v_i,w_i$, the Hilbert space of the system will decouple into a low-energy subspace, with zero occupation in the middle sites, and a high-energy subspace. As indicated in \cite{Mukherjee2018}, the low-energy subspace has the geometry of a Creutz ladder. Hence, changing the tunneling amplitudes in certain ways will let us obtain the different CSSH ladders. In particular, all topological flat band ladders and the SSH-like topological zero modes of the $\hgls$-CSSH ladder can be implemented in a straight-forward way. This can allow experimentalists to explore the various topological zero modes of Section \ref{sec:topo} and Appendix \ref{ap:edge}, and the flat-band dynamics of the models. The hopping amplitudes can be tuned by simply changing the distance between waveguides.

The low-energy effective Hamiltonian, obtained using a Schrieffer-Wolff transformation \cite{Hofmann2012}, is:
\begin{align}
    \sH_{\eff} = \sH_{C\!S\!S\!H} + \sum_{n,\alpha} \epsilon_\alpha \kb{n,\alpha}{n,\alpha}, \label{eq:Heff}
\end{align}
where $\alpha=A,B,C,D$ and $\sH_{C\!S\!S\!H}$ is the CSSH Hamiltonian of equation \eqref{eq:CSSH}, with the following parameters:
\begin{align}
    &J_1^+ = \frac{v_3 w_1}{\eps}e^{i\phi/2}; \hspace{20pt}
     J_2^+ = \frac{w_2 v_4}{\eps}e^{i\phi/2}\nonumber\\
     &J_3^+ = \frac{w_3 v_1}{\eps}e^{i\phi/2}; \hspace{20pt}
     J_4^+ = \frac{v_2 w_4}{\eps}e^{i\phi/2}\nonumber\\
     &t_1 = \frac{v_3 w_2}{\eps}; \hspace{20pt}
     t_2 = \frac{w_1 v_4}{\eps}\nonumber\\
     &t_3 = \frac{w_3 v_2}{\eps}; \hspace{20pt}
     t_4 = \frac{v_1 w_4}{\eps}\nonumber\\
    &m_1 = \frac{1}{\eps}(v_2v_1 e^{i\phi/2} + v_3v_4 e^{-i\phi/2})\nonumber\\
    &m_2 = \frac{1}{\eps}(w_2w_1 e^{i\phi/2} + w_3w_4 e^{-i\phi/2})\nonumber\\
    &\eps_A = \frac{v_1^2 + v_3^2}{\eps};\hspace{20pt} \eps_B = \frac{v_2^2 + v_4^2}{\eps}\nonumber\\
    &\eps_C = \frac{w_1^2 + w_3^2}{\eps};\hspace{20pt} \eps_D = \frac{w_2^2 + w_4^2}{\eps}.\nonumber
\end{align}
In the most general case, the vertical amplitudes are not real.

The hourglass CSSH ladder can be obtained by choosing the parameters:
\begin{align}
    &v_1=v_2=w_3=w_4=s'\nonumber\\
    &v_3=v_4=w_1=w_2=s,\nonumber
\end{align}
which generates the amplitudes:
\begin{align}
    &J=\frac{s^2}{\eps};\hspace{20pt} J'=\frac{s'^2}{\eps}\nonumber\\ &m_1=\frac{1}{\eps}(s'^2e^{i\phi/2} + s^2 e^{-i\phi/2})=m_2^*,\nonumber
\end{align}
with an additional on-site energy of $(s^2+s'^2)/\eps$ on all sites. This model is represented in Fig. \ref{fig:rhombi} (b), with long (short) links to represent weak (strong) tunneling amplitudes.
If $\phi=\pi$, $m_1=m_2=0$ and the flat-band $\hgls$-CSSH ladder is obtained. If $\phi=0$, $m_1=m_2=(s^2+s'^2)/\eps$ and the ladder has SSH-like topological zero modes if $s'<s$. For other values of $\phi$, the vertical links have the same magnitude and opposite phase. They are always non-zero, so the rungless, non-$\pi$ flux topological case cannot be obtained in this way.

The rhomboid CSSH ladder Hamiltonian is obtained from \eqref{eq:Heff} if we choose:
\begin{align}
    v_2=v_3=s';\hspace{20pt} v_1=v_4=s; \hspace{20pt} w_i=w\,\,\,\forall i,\nonumber
\end{align}
with the parameters:
\begin{align}
    &J=\frac{sw}{\eps};\hspace{20pt} J'=\frac{s'w}{\eps}\nonumber\\ &m_1=\frac{2ss'}{\eps}\cos{\phi/2};\hspace{10pt} m_2=\frac{2w^2}{\eps}\cos{\phi/2}\nonumber\\ &\eps_A=\eps_B=\frac{s^2+s'^2}{\eps};\hspace{20pt} \eps_C=\eps_D= \frac{2w^2}{\eps}\nonumber
\end{align}
This lattice is depicted in Fig. \ref{fig:rhombi} (c). In the $\pi$-flux case, $m_1=m_2=0$, and we can choose $w=\sqrt{s^2+s'^2}$ to keep all on-site energies $\eps_\alpha$ equal and obtain the topological flat-band $\rhomb$-CSSH ladder. For other values of $\phi$, $w=\sqrt{ss'}$ might be chosen to get $m_1=m_2$, and then the on-site energies $\eps_\alpha$ could be compensated using the appropriate chemical potentials on the rhombus chain sites. Like in the hourglass case, the rungless non-$\pi$-flux case cannot be achieved in this way.

Finally, to implement the delta CSSH ladder, we set:
\begin{align}
    v_1=v_3=s';\hspace{15pt} v_2=v_4=s; \hspace{15pt} w_i=\sqrt{ss'}\,\,\,\forall i,\nonumber
\end{align}
and obtain [see Fig. \ref{fig:rhombi} (d)]:
\begin{align}
    &J=\frac{s^2s'}{\eps};\hspace{20pt} J'=\frac{s'^2s}{\eps}\nonumber\\ &m_1=m_2=m=\frac{2ss'}{\eps}\cos{\phi/2}; \hspace{20pt} \eps_A=\frac{2s'^2}{\eps}\nonumber\\
    &\eps_B=\frac{2s^2}{\eps};\hspace{20pt} \eps_C=\eps_D= \frac{2ss'}{\eps}.\nonumber
\end{align}

In this model, the energies $\eps_\alpha$ cannot be made equal by fixing $w$, so they must be compensated using on-site potentials.
If $\phi=\pi$, then $m=0$ and the topological flat-band case is obtained. Like in the other cases, the $\phi\ne\pi$ regime is not a strong topological phase, because $m$ is finite there. The $\pi$-flux case with $m\ne 0$, which belongs to the D class, cannot be obtained directly using this method. To implement it, an additional tunneling amplitude of value $m$ could be established between the top and bottom sites of each rhombi, to create the vertical links of the ladder. This makes the lattice similar to the cold atom proposal in \cite{Sun2017}, and both of them could be used to implement the rungless topological regime for any value of $\phi$, and also the $\pi$-flux with $m\ne 0$ case. In the $\Delta$-CSSH ladder, the latter corresponds to the D-class TI.

\providecommand*\hyphen{-}

\end{document}